\documentclass{ieeeaccess}
\usepackage{cite}
\usepackage{amsmath,amssymb,amsfonts}
\usepackage{graphicx}
\usepackage{textcomp}
\usepackage{url}
\usepackage[ruled]{algorithm2e}
\usepackage[noend]{algpseudocode}
\usepackage{xcolor}
\usepackage{booktabs}
\newcommand{\rev}[1]
{{\color{black}#1}}

\DeclareMathOperator{\E}{E}
\newcommand{\K}[0]{\mathcal{K}}

\newcommand{\ad}[0]{{\rm ad}}
\newcommand{\Diff}[0]{{\rm Diff}}

\newcommand{\Dist}[0]{{\rm Dist}}

\DeclareMathOperator{\Slerp}{Slerp}
\DeclareMathOperator{\clip}{clip}

\usepackage{bm}
\makeatletter
\AtBeginDocument{\DeclareMathVersion{bold}
\SetSymbolFont{operators}{bold}{T1}{times}{b}{n}
\SetSymbolFont{NewLetters}{bold}{T1}{times}{b}{it}
\SetMathAlphabet{\mathrm}{bold}{T1}{times}{b}{n}
\SetMathAlphabet{\mathit}{bold}{T1}{times}{b}{it}
\SetMathAlphabet{\mathbf}{bold}{T1}{times}{b}{n}
\SetMathAlphabet{\mathtt}{bold}{OT1}{pcr}{b}{n}
\SetSymbolFont{symbols}{bold}{OMS}{cmsy}{b}{n}
\renewcommand\boldmath{\@nomath\boldmath\mathversion{bold}}}
\makeatother

\def\BibTeX{{\rm B\kern-.05em{\sc i\kern-.025em b}\kern-.08em
    T\kern-.1667em\lower.7ex\hbox{E}\kern-.125emX}}

\begin{document}
\history{Date of publication xxxx 00, 0000, date of current version xxxx 00, 0000.}
\doi{10.1109/ACCESS.2026.xxxxxxx}

\title{A Unified Deep Learning Framework for Motion Correction in Medical Imaging}
\author{\uppercase{Jian Wang}\authorrefmark{1},
\uppercase{Razieh Faghihpirayesh}\authorrefmark{2,1}, \uppercase{Danny Joca}\authorrefmark{1}, \uppercase{Polina Golland}\authorrefmark{3},  AND \uppercase{Ali Gholipour}\authorrefmark{4,5,6,1},
\IEEEmembership{Senior Member, IEEE}}

\address[1]{Department of Radiology, Boston Children's Hospital, and Harvard Medical School, Boston, MA 02115 USA}
\address[2]{Department of Electrical and Computer Engineering, Northeastern University, Boston, MA 02115 USA}
\address[3]{Department of Electrical Engineering and Computer Science, Massachusetts Institute of Technology, Cambridge, MA 02139 USA}
\address[4]{Department of Radiological Sciences, University of California Irvine, Irvine, CA 92697 USA}
\address[5]{Department of Electrical Engineering and Computer Science, University of California Irvine, Irvine, CA 92697 USA}
\address[6]{Department of Biomedical Engineering, University of California Irvine, Irvine, CA 92697 USA}
\tfootnote{This research was supported in part by the National Institutes of Health (NIH) under award numbers R01EB031849, R01EB032366, R01HD109395 and R01EB032708; in part by the Office of the Director of the NIH under award number S10OD025111, and in part by NVIDIA Corporation. The content of this publication is solely the responsibility of the authors and does not necessarily represent the official views of the NIH or NVIDIA.}

\markboth
{Wang \headeretal: A Unified Deep Learning Framework for Motion Correction in Medical Imaging}
{Wang \headeretal: A Unified Deep Learning Framework for Motion Correction in Medical Imaging}

\corresp{Corresponding author: Ali Gholipour (e-mail: ali.gholipour@uci.edu).}

\begin{abstract}
Deep learning has shown significant value in medical image registration for motion correction, however, current techniques are either limited by the type and range of motion they can handle, or require iterative inference and/or retraining for new imaging data. To address these limitations, we introduce UniMo, a Unified Motion Correction framework that leverages deep neural networks to correct for various types of motion in medical imaging. UniMo exploits an alternating optimization scheme for a unified loss function to train an integrated model of 1) an equivariant neural network for global rigid motion correction and 2) an encoder-decoder network to correct local deformations. To this end, it features a geometric deformation augmenter that 1) enhances the robustness of global motion correction by addressing any local deformations whether they are caused by non-rigid motion or geometric distortions, and 2) generates augmented data to improve the training process. UniMo is a hybrid model that uses both image intensities and shapes to achieve robust performance amid image appearance variations, and, therefore, it generalizes well to various medical imaging modalities without a need for network retraining. We trained and tested UniMo to track motion in fetal magnetic resonance imaging, which is a very challenging application due to 1) the presence of both large rigid and non-rigid motion, and 2) large variations in image appearance. Then we tested the trained model, without retraining, on various image modalities from three public datasets, including MedMNIST, lung CT, and BraTS. The results show that UniMo surpassed existing motion correction methods in terms of accuracy, and, notably, it enabled one-time training on a single modality while maintaining high stability and adaptability for inference across multiple unseen imaging datasets.
By offering a unified solution to motion correction, UniMo marks a significant advantage in challenging applications with a mixture of bulk motion and local deformations. The code is available at: \url{https://github.com/IntelligentImaging/UNIMO}
\end{abstract}

\begin{keywords}
Deep Learning, Medical Image Registration, Motion Correction, Rigid Registration, Deformable Registration, Multi-Feature Fusion, Equivariant Neural Network
\end{keywords}

\titlepgskip=-21pt

\maketitle

\section{Introduction}
\label{sec:introduction}

%\subsection{Motion in Medical Imaging}

\PARstart{M}{otion} remains a significant challenge in medical imaging. Depending on the modality and source, it can cause artifacts such as ghosting, blurring, smearing, streaking, and signal loss often rendering images unreliable or clinically unusable~\cite{zaitsev2015motion,maclaren2013prospective,kyme2021motion,silva2018challenges,runge2019motion,wallace2021free}. Standard practice relies on prevention and mitigation strategies, including patient training, breath-holding, and motion-robust acquisition techniques~\cite{zaitsev2015motion}. However, these approaches are not always feasible; involuntary motion—such as fetal or cardiac motion—cannot be prevented, and cooperation is often limited in pediatric or high-acuity patients~\cite{malamateniou2013motion,jaimes2016strategies,harrington2022strategies,uus2023retrospective}.

The impact of motion is especially pronounced in slow modalities like magnetic resonance imaging (MRI), where long acquisition times increase the likelihood of artifacts%Despite substantial progress, it remains a complex issue with only partial solutions.
~\cite{zaitsev2015motion}.
An ideal starting point is a fast, robust acquisition strategy that minimizes the likelihood and impact of motion while enabling its detection and correction. Ultra-fast methods such as advanced echo-planar imaging (EPI) techniques achieve high spatiotemporal resolution for 3D and 4D imaging~\cite{wang2019echo,dong2025romer,dai2025instantaneous,yang2026rapid,ge2026high}. Techniques like parallel imaging~\cite{deshmane2012parallel} and compressed sensing~\cite{lustig2007sparse} accelerate acquisition through under-sampling. Non-Cartesian sampling strategies like PROPELLER~\cite{pipe1999motion}, radial, stack-of-stars, kooshball~\cite{jung2009k,usman2013motion,feng2014golden,christodoulou2018magnetic,feng2022golden,wallace2024rapid,schauman2026exploration}, and rosette~\cite{noll1997multishot} are of interest since they provide dense sampling of the K-space center, which is useful to estimate and correct motion. These fast, motion-robust strategies are often combined with prospective (real-time) or retrospective (offline) processing for motion-corrected image reconstruction.
%While these fast, motion-robust strategies reduce artifacts, they often require additional processing steps for motion correction.}

While cardiac and respiratory gating can synchronize imaging with physiological cycles~\cite{ehman1984magnetic,larson2004self,di2019automated,ludwig2021pilot,anand2024beat}, motion correction is generally needed to image moving subjects or organs~\cite{usman2013motion,moghari2013free}. This process relies on accurate motion measurement or estimation, which can be performed prospectively~\cite{thesen2000prospective,welch2002spherical,van2006real,tisdall2012volumetric,white2010promo,maclaren2013prospective,zaitsev2015motion,zaitsev2017prospective,slipsager2019markerless,slipsager2021comparison,adanyeguh2024fast,jayadev2025accelerated}, retrospectively~\cite{hansen2012retrospective,parkes2018evaluation,gallichan2016retrospective,marami2016motion,uus2023retrospective,godenschweger2016motion,isaieva2025optimal}, or via hybrid approaches~\cite{aksoy2012hybrid,maclaren2011combined,gholipour2011motion}. The primary performance metrics are spatial accuracy and temporal resolution; however, clinical adoption is often dictated by a method's complexity, cost, and physical requirements~\cite{zaitsev2015motion,andronesi2021motion}. For instance, camera-based tracking~\cite{olesen2012list,slipsager2019markerless} offers high spatiotemporal resolution, which is great, but it requires specialized hardware and setup and cannot be used for internal targets like fetal motion.

Internal motion navigators are more prevalent than external motion navigators. While rigid-body motion can be inferred directly from the complex-valued MRI signal or from free-induction decay signals measured at the 
K-space center~\cite{kober2011head,wallace2024rapid}, most internal motion navigation techniques—such as volume (vNAV)~\cite{van2006real,tisdall2012volumetric}, spiral (sNAV)~\cite{welch2002spherical,adanyeguh2024fast} and fat (FatNav) navigators~\cite{gallichan2016retrospective}—rely on image registration. Furthermore, image registration serves as the foundation for the vast majority of retrospective motion correction algorithms~\cite{hansen2012retrospective,gholipour2010robust,godenschweger2016motion,parkes2018evaluation,gallichan2016retrospective,kainz2015fast,marami2016motion,alansary2017pvr,uus2020deformable,uus2023retrospective} as well as the majority of prospective image-based (including self-navigated) motion tracking systems~\cite{thesen2000prospective,welch2002spherical,van2006real,tisdall2012volumetric,white2010promo,zaitsev2017prospective,slipsager2021comparison,dosenbach2017real,sui2020slimm,marchetto2023robust,adanyeguh2024fast,neves2023real,verdera2025heron,jayadev2025accelerated,marchetto2026contrast}.

Therefore, image registration plays a critical role in motion correction in medical imaging. Although classical image registration methods can be effective for motion correction, they often necessitate full execution of an optimization process on non-convex cost functions that only act as surrogate measures of image alignment~\cite{fitzpatrick2000image,white2010promo,ferrante2017slice,gholipour2019biomedical,sui2020slimm}. Thus, these techniques are typically limited by a narrow capture range, especially when addressing large motions~\cite{salehi2018real}. This constraint reduces their precision and applicability in challenging applications such as fetal MRI where subjects may move in wide angles~\cite{salehi2018real,singh2020deep,calixto2024advances}. To address these limitations, recent research has increasingly turned to deep learning based models for motion estimation/correction~\cite{spieker2023deep}.

\subsection{Deep Learning for Motion Correction}

%affecting areas such as image segmentation~\cite{faghihpirayesh2022deep}, reconstruction~\cite{xu2022svort}, and pose estimation~\cite{golland20203d}. Traditional motion correction techniques~\cite{gholipour2010robust,kuklisova2012reconstruction,malamateniou2013motion} have undergone substantial evolution over time. 
Deep learning-based motion correction techniques have seen widespread applications across various imaging modalities, with significant developments in two main areas:  i) A major focus on MRI, including structural, functional, diffusion, and quantitative MRI. These models are extensively studied due to their broad applicability and substantial impact in the field, covering a wide range of clinical scenarios.  ii) Motion correction for other modalities, such as ultrasound, positron emission tomography (PET), and Computed Tomography (CT) where various advancements are also being explored.

Learning-based motion correction in MRI focuses on two key approaches: i) image-based motion correction and ii) techniques using k-space data. Image-based methods target motion correction by processing reconstructed MRI images with deep neural networks, such as Convolutional Neural Networks (CNNs)~\cite{salehi2018real,evan2022keymorph} and other deep learning architectures~\cite{wang2024joint,wang2024spaer,moyer2021equivariant,xu2022svort}. These approaches often employ convolutional encoder-decoder networks, leveraging downsampling and upsampling for feature extraction and reconstruction~\cite{haskell2019network,schlemper2017deep}. Advanced models like U-Nets, recurrent neural networks, and transformers have been proposed to handle temporal dependencies and spatial relationships~\cite{singh2020deep,lyu2021cine,xu2022svort,xu2023nesvor,wang2024spaer}. Additionally, prior-assisted methods improve correction by incorporating supplementary data such as multi-contrast or dynamic images, enhancing performance through architectures that integrate this information~\cite{chatterjee2020retrospective,al2023knowledge}. These methods offer robust solutions by combining motion correction with other learning-based tasks, ultimately improving MRI reconstruction quality, e.g.~\cite{xu2023nesvor}.

K-space-based motion correction methods, on the other hand, leverage raw k-space data with MR reconstruction to mitigate motion artifacts through deep learning techniques. These include deep neural networks for noise-resilient motion correction and motion parameter estimation~\cite{kuzmina2022autofocusing+,hossbach2023deep,eichhorn2024physics}, and methods that combine motion detection with traditional reconstruction~\cite{singh2022joint}. One approach targets artifacts by detecting motion directly from k-space images and data using CNNs~\cite{cui2023motion}, while others use temporal information~\cite{seegoolam2019exploiting} or end-to-end motion models to enhance reconstruction quality and reduce acquisition times~\cite{qi2021end}. Additionally, generative models and implicit neural representations (INRs) offer subject-specific solutions to handle unique motion characteristics~\cite{huang2023neural,spieker2023iconik}.

In the domain of ultrasound and PET imaging also, significant advancements in motion correction have been achieved. For instance, approaches~\cite{harput2018two} that combine affine and nonrigid motion estimation have proven effective in addressing both large-scale and fine deformations, leading to improved image quality. The use of CNNs for 3D freehand ultrasound reconstruction has streamlined the motion estimation process~\cite{prevost20183d}, bypassing traditional physical models, while resulting in more accurate and reliable ultrasound imaging. Motion correction techniques based on PET utilizing bidirectional 3D long short-term memory networks have been effective in addressing inter-frame rigid motion in dynamic cardiac PET, improving both motion estimation and myocardial blood flow quantification~\cite{shi2021automatic}. Similarly, deep learning models designed to predict rigid motion parameters directly from PET data have shown promise in correcting head motion in brain PET scans~\cite{li2021deep}. 

Despite significant advancements, state-of-the-art methods remain heavily constrained by the specific imaging datasets and modalities they are trained on and the types of motion they can handle, which limit their effectiveness in correcting motion in unseen imaging domains. While one approach~\cite{moyer2021equivariant} shows promise in handling motion for unseen modalities, it struggles with large motions in image pairs that are heavily contaminated by noise~\cite{billot2024se} and those that exhibit contrast changes.
For instance, substantial intensity and contrast variations in images pose significant challenges, as current algorithms largely rely on image intensities while overlooking shape information. This limits generalization capacity of the trained models.

On the other hand, continuous advances in deep learning have enabled reliable and accurate segmentation and landmark detection, supporting their effective use in image registration. Compared to traditional intensity-based methods, shape- and keypoint-based approaches~\cite{evan2022keymorph,wang2023robust,wang2023metamorph,wang2024brainmorph} now provide robust alternatives, particularly in challenging imaging scenarios. Building on these developments, our joint optimization framework integrates shape cues to improve registration performance. This reflects a broader shift in the field toward leveraging geometric representations derived from learned features, which have demonstrated strong generalization across datasets and modalities~\cite{wang2023robust}.
Motivated by the need for generalizable and accurate motion correction across diverse imaging conditions, we introduce UniMo---a unified method that combines shape and intensity information to correct both bulk rigid motion and local deformations. UniMo capitalizes on the strengths of deep learning while avoiding the need for extensive retraining. Therefore, it offers a robust and efficient solution that is adaptable to various imaging conditions. By not imposing restrictive assumptions about motion types or ranges, UniMo ensures stable and accurate correction for a range of applications. Moreover, eliminating the retraining requirement reduces computational overhead and enables seamless integration into imaging workflows, which advances the state of the art in motion estimation and correction.

This paper is a major extension over our prior conference work. In prior work~\cite{wang2024joint} we introduced joint rigid and deformable registration for single-modality fetal brain motion correction using image intensities only. In~\cite{wang2024spaer} we extended rigid motion correction to spatio-temporal motion tracking in 4D fetal EPI. UniMo is a substantial extension of those prior works as it includes: (1) a new hybrid image-plus-shape learning framework with SLERP (\rev{spherical} linear interpolation)-based rotation fusion on $SO(3)$ that enables cross-modality generalization without retraining, (2) comprehensive cross-modality experiments on four distinct imaging datasets, and (3) a detailed motion simulation protocol and discussion of importance and practical value of the proposed framework.

The main contributions of this work (UniMo) are: 
\begin{itemize}
    \item UniMo is the first method for generalizable correction of both rigid and non-rigid motion without a need for retraining in a new domain. 
    \item UniMo offers motion correction of a pair of images with stable convergence in training and real-time inference. 
    \item  UniMo capitalizes on the synergies between global rigid motion estimation and local deformation correction. This innovative strategy demonstrates its versatility through its broad applicability to various joint learning tasks.
\end{itemize}

\section{Background}
In this section, we review the theories of rigid motion correction, deformable image registration, and joint optimization.

\subsection{Rigid Motion Estimation via Equivariant Filters}
Rigid motion estimation aims to identify the best translation $\mathcal{T}$ and rotation $\mathcal{R}$ parameters that define a rigid transformation $Q(\mathcal{T},\mathcal{R})$ between a pair of images. In a simple form, this process may be formulated as minimizing the Euclidean distance $d$ between a source image $S$ and a target image $T$,
\begin{equation}
\label{eq:imRigid} 
\E [Q (\mathcal{T},\mathcal{R})] =  \text{Dist} [S \circ Q (\mathcal{T},\mathcal{R}), T] + \text{Reg}(\mathcal{T},\mathcal{R}) ,
\end{equation} 
where $\circ$ is the composition operator that resamples $S$ using the rigid transformation $Q(\mathcal{T},\mathcal{R})$. When this operator is applied to any vector $\textbf{v}$, it yields a transformed vector $Q(\textbf{v})$ of the form $Q(\textbf{v}) = \mathcal{R} \textbf{v} + \mathcal{T}$. Here, $\mathcal{R}^T = \mathcal{R}^{-1}$ indicating that $\mathcal{R}$ is an orthogonal matrix. 

Rather than estimating the transformation function $Q$ directly in the original image space, efficient approaches have demonstrated that it can be computed using low-dimensional representations, such as equivariant spatial means or equivariant features of images~\cite{moyer2021equivariant}. In specific, equivariant filters have shown proven stability and accuracy in rigid image registration. The closed-form solution for both translation and rotation parameters is:
\begin{equation}
\label{eq:closeform} 
\mathcal{T} = \bar{T} - \mathcal{R}\bar{S},\quad \mathcal {R} = V \cdot {U}^T,\quad \text{s.t.} \, \, \det (\mathcal{R}) = 1,
\end{equation} 
where $\bar{S}$, $\bar{T}$ represent the low-dimensional representations of the source and target images respectively. Specifically, $\bar{S}$ and $\bar{T}$ are $K$-dimensional vectors ($K=128$ in our implementation) obtained by computing the spatial mean of each equivariant filter response over all voxel locations. These vectors preserve equivariance under rigid transformations, enabling the following closed-form SVD solution. ${U} \Sigma V^* = \bar{S} \cdot \bar{T}^{T}$, $U$ and $V^*$ are real orthogonal matrices, $\Sigma$ is a diagonal matrix with non-negative real numbers on the diagonal. The condition on the determinant of $\mathcal{R}$ in Eq.~\eqref{eq:closeform} is set to ensure it accurately reflects a rigid rotation.

\subsection{Deformation Correction via Deformable Registration}
In addition to global rigid transformations, deformations may be needed to align medical images to compensate for elasticity of tissue/organs, anatomical variations (between subjects or across time), or geometric distortions as imaging artifacts. In this section, we provide an overview of the Large Deformation Diffeomorphic Metric Mapping (LDDMM) algorithm for image registration~\cite{beg2005computing} between the rigid motion-corrected image $S \circ Q (\mathcal{T},\mathcal{R})$ and the target image $T$.
For simplicity, we denote $S \circ  Q (\mathcal{T},\mathcal{R})$ as $\mathbf{\hat{S}}$.

Let $\mathbf{\hat{S}}$ be a source image and $T$ be a target image defined on a torus domain
$\Gamma = \mathbb{R}^d / \mathbb{Z}^d$ ($\mathbf{\hat{S}}(x), T(x) : \Gamma \rightarrow \mathbb{R}$). 
The problem of diffeomorphic image registration is to find the shortest path to generate time-varying diffeomorphisms
$\{\psi_t\}: t \in [0,1] $ such that $\mathbf{\hat{S}} \circ \psi_1$ is similar to $T$, where $\circ$ is a composition operator that resamples $\mathbf{\hat{S}}$
by the smooth mapping $\psi_1$. This is typically solved by minimizing the energy function of LDDMM~\cite{beg2005computing} over an initial velocity field $v_0$. 

For computational efficiency, we employ a fast version of LDDMM that characterizes deformations as $\{\psi_t\}$ in a low-dimensional band limited space. The corresponding time-dependent tangent vector of such deformations can be determined by an initial condition $\tilde{v}_0$, which is the initial velocity field represented in the bandlimited Fourier domain (the tilde notation denotes Fourier-domain quantities throughout), 
\begin{align}
\label{eq:LDDMM} 
\text{E}(\tilde{v}_0) = \text{Dist} (\mathbf{\hat{S}} \circ \psi_1, T) + (\tilde{\mathcal{L}} \tilde{v}_0, \tilde{v}_0), 
\end{align} 
where the distance function $\Dist(\cdot , \cdot)$
measures the dissimilarity between images. Commonly used distance functions include sum-of-squared difference
of image intensities~\cite{beg2005computing}, normalized cross correlation~\cite{avants2008symmetric}, and mutual
information~\cite{wells1996multi,wang2023metamorph}. The regularization term $(\tilde{\mathcal{L}} \tilde{v}_0, \tilde{v}_0)$ enforces spatial 
smoothness of transformations with $\tilde{\mathcal{L}}$ being a symmetric and positive-definite differential operator in the Fourier domain. It converts a vector field $\tilde{v}$ to a momentum vector by $\tilde{m} = \tilde{\mathcal{L}}\tilde{v}$. Here $(\cdot, \cdot)$ denotes an $L^2$ inner product weighted by $\tilde{\mathcal{L}}$; intuitively, minimizing this term encourages smoother, more regular deformation fields. The deformation $\psi_{1}$ corresponds to $\tilde{\psi}_1$ in Fourier space via the Fourier transform $\mathcal{F}(\psi_{1})=\tilde{\psi}_1$, or its inverse $\psi_{1}=\mathcal{F}^{-1}(\tilde{\psi}_1)$.

Let $\widetilde{\Diff}(\Omega)$ and $\tilde{V}$ denote the bandlimited space of diffeomorphisms and velocity fields respectively. The Euler-Poincar\'e differential (EPDiff) equation~\cite{arnold1966,miller2006geodesic} is reformulated in a complex-valued Fourier space with much less dimensions, i.e., 
\begin{align}\label{eq:epdifffourier}
    \frac{\partial \tilde{v}_t}{\partial t} =-\tilde{\K}\left[(\tilde{\mathcal{D}} \tilde{v}_t)^T \star \tilde{\mathcal{L}}\tilde{v}_t + \tilde{\nabla} \cdot (\tilde{\mathcal{L}}\tilde{v}_t \otimes \tilde{v}_t) \right],
\end{align}
where $\star$ is a circular matrix-vector field auto-correlation\footnote{Auto-correlation operates on zero-padded signals followed by truncating to the bandlimits in each dimension to
ensure the output remains bandlimited.}. $\tilde{\mathcal{K}}$ is a smoothing operator and $\tilde{\mathcal{D}}\tilde{v}$ is a tensor product with $\tilde{\mathcal{D}} (\mathbf{\xi}) = i \sin(2 \pi \mathbf{\xi})$ representing the Fourier frequencies of a central difference Jacobian matrix $D$. The operator $\tilde{\nabla} \cdot$ is the discrete divergence operator that is computed as the sum of the Fourier coefficients of the central difference operator $\tilde{D}$ along each dimension, i.e., $\tilde{\nabla} \cdot \mathbf{\xi} = \sum\limits_{j=1}^d i \sin(2 \pi \mathbf{\xi}_j)$. Since $\tilde{\mathcal{K}}$ on the left side of Eq.~\eqref{eq:epdifffourier} is a low-pass filter that suppresses high frequencies in the Fourier domain, all operations are easy to implement in a truncated low-dimensional space by eliminating high frequencies. 

The diffeomorphic transformations can also be represented in the frequency domain~\cite{zhang2019fast} as $\tilde{\psi}_t$, 
\begin{align}
\label{eq:lowtransformation}
\frac{d \tilde{\psi}_t }{dt}&=-\tilde{\mathcal{D}} \tilde{\psi}_t \ast \tilde{v}_t, \quad 
\end{align}
where $*$ is a circular convolution. 

\subsection{Joint Optimization}
A joint optimization approach estimates both rigid transformation $Q(\mathcal{T}, \mathcal{R})$ and deformation $\tilde{v}_0$ by integrating a combination of Eq.~\eqref{eq:imRigid} and Eq.~\eqref{eq:LDDMM},
\begin{align}
\label{eq:jointobj} 
\E[Q (\mathcal{T},\mathcal{R}),\tilde{v}_0] &= \E [Q (\mathcal{T},\mathcal{R})] + \text{E}(\tilde{v}_0) \nonumber\\
& =  \underbrace {\text{Dist} [S \circ Q (\mathcal{T},\mathcal{R}), T] + \text{Reg}(\mathcal{T},\mathcal{R})}_{\text{Rigid Motion Correction}} \nonumber
\\ 
& +\underbrace {\text{Dist} (\mathbf{\hat{S}} \circ \psi_1 (\tilde{v}_0)), T) + (\tilde{\mathcal{L}} \tilde{v}_0, \tilde{v}_0)}_{\text{Deformation Correction}}, \nonumber \\ &\, \,  \underbrace {s.t. \, \, \text{Eq.}~\eqref{eq:closeform} ,\, \,\text{Eq.}~\eqref{eq:epdifffourier}\ \& \ \text{Eq.}~\eqref{eq:lowtransformation}}_{\text{Joint Constraints}}
\end{align}
Such joint approaches have demonstrated enhanced accuracy and robustness for single image modalities~\cite{wang2024joint,wang2024spaer}. However, these frameworks fall short when image pairs exhibit significant intensity changes.
The issue arises because both dissimilarity terms in Eq.\eqref{eq:jointobj} are highly dependent on the images themselves, increasing the likelihood that the optimization will be biased by intensity variations. This limitation motivated the development of a hybrid motion correction framework that integrates knowledge from both image intensities and shape to effectively reduce errors caused by image intensity variations.

\section{Methodology}
\subsection{Hybrid Motion Correction}
In this section, we present a hybrid motion correction approach that utilizes both image and shape information. UniMo requires both images and their corresponding segmentation maps as input during training and inference. Segmentation maps are converted to distance transforms, which serve as intensity-invariant shape representations for the geometric branch of the model. In practice, segmentations can be obtained automatically using existing deep learning methods (e.g.,~\cite{faghihpirayesh2024fetal} for fetal brains) or from public dataset annotations. Our objective is to develop an optimal motion correction solution, denoted as $\pmb{Q} (\pmb{\mathcal{T}},\pmb{\mathcal{R}})$, which exhibits increased robustness to variations in image intensity. \rev{Let $Q_I(\mathcal{T}_I,\mathcal{R}_I)$ and $Q_G(\mathcal{T}_G,\mathcal{R}_G)$ denote the rigid transformations estimated from images and segmentations separately. We define the fused rigid transformation $\pmb{Q}(\pmb{\mathcal{T}},\pmb{\mathcal{R}})$ by}
\begin{align}
\rev{\pmb{\mathcal{R}}} &\rev{= \Slerp(\mathcal{R}_I, \mathcal{R}_G;\, \lambda),} \nonumber \\
\label{eq: mmq}
\rev{\pmb{\mathcal{T}}} &\rev{= (1-\lambda)\,\mathcal{T}_I + \lambda\,\mathcal{T}_G, \qquad \lambda \in [0,1],}
\end{align}
\rev{where $\Slerp(\cdot,\cdot;\lambda)$ denotes spherical linear interpolation between the two rotations along the geodesic on $SO(3)$ (Section~III-B), and $\lambda$ is the weight parameter balancing the effect of both domains. Setting $\lambda = 0$ recovers the image-only estimate and $\lambda = 1$ recovers the shape-only estimate. Rotation matrices are not combined by weighted addition, because a weighted sum of two orthogonal matrices is in general not orthogonal; the rotation is therefore fused on the manifold, and only the translation vectors, which belong to a vector space, are combined linearly.} In this work, we apply distance transform to segmentation maps to represent the most straightforward geometric shape of images. Other intensity-invariant shape descriptors~\cite{vranic20013d,khotanzad1990invariant} can be easily integrated into our framework as well.

After obtaining the hybrid rigid transformation $\pmb{Q}$, for deformation estimation we estimate $\tilde{v}^I_0$ and $\tilde{v}^G_0$ (representing the transformation fields) in the spatial domain for images and segmentations respectively. The derivation of a hybrid velocity field is not the primary objective of this work. Instead, we formalize the hybrid objective function for optimization as follows,
\begin{align}
\label{eq:total} 
\E&[\pmb{Q},  \tilde{v}^I_0,\tilde{v}^G_0] = \E[\pmb{Q}(\pmb{\mathcal{T}},\pmb{\mathcal{R})}] + \E(\tilde{v}^I_0) + \E(\tilde{v}^G_0).
\end{align}

\subsection{Hybrid Rigid Transformation Fusion By SLERP}
In this section, we provide a step-by-step derivation for computing $\pmb{Q}$. To leverage the complementary strengths of multiple modalities, we propose a hybrid image-plus-shape fusion framework. 
Within this framework, rotations are estimated independently from distinct modalities; however, to achieve a unified transformation, these estimates must be integrated effectively. Unlike traditional methods that rely on simple linear addition, which can lead to non-orthogonal results, we employ Spherical Linear Interpolation (SLERP) to fuse the rotations~\cite{jafari2014spherical}. This approach ensures that the resulting transformation $\pmb{Q}$ strictly maintains the properties of the $SO(3)$ manifold. By following the shortest path, or geodesic, on the unit sphere, SLERP enables smooth and consistent interpolation between disparate rotational estimates while preserving the essential orthogonality and unit length of the transformations. 
By avoiding the pitfalls of directly averaging rotation matrices or quaternions, our method ensures that the combined rotation remains both mathematically rigorous and physically meaningful. 
This preservation of the rotational manifold's integrity is crucial for cross-modality generalization in medical imaging, where reliable motion correction is required across diverse datasets without retraining~\cite{zhong2024slerpface,jang2024spherical,li2022quatse}.

We derive $\pmb{Q}$ by computing two rigid transformation matrices, 
\[
\mathcal{Q}_I = \begin{bmatrix}
\mathcal{R}_I & \mathbf{t}_I \\
\mathbf{0}^T & 1
\end{bmatrix},
\quad \mathcal{Q}_G = \begin{bmatrix}
\mathcal{R}_G & \mathbf{t}_G \\
\mathbf{0}^T & 1
\end{bmatrix},
\]

\begin{enumerate}
    \item  Convert \( \mathcal{R}_I \) and \( \mathcal{R}_G \) to quaternions \( \mathbf{q}_I \) and \( \mathbf{q}_G \);
    \item Apply spherical linear interpolation to \( \mathbf{q}_I \) and \( \mathbf{q}_G \) with weight \( \lambda \rev{\in [0,1]} \) and \rev{an angle} $\theta$ \rev{between the two quaternions} (Sec. A in supplementary),
    \[
    \mathbf{q} = \frac{\sin((1 - \lambda) \theta)}{\sin(\theta)} \cdot \mathbf{q}_I + \frac{\sin(\lambda \theta)}{\sin(\theta)} \cdot \mathbf{q}_G
    ;\]
    \item Convert \( \mathbf{q}\) back to a rotation matrix \( \pmb{\mathcal{R}} \);
    \item Linearly combine translations,
    \[
    \pmb{\mathcal{T}} = (1-\lambda)\mathcal{T}_I +\lambda \mathcal{T}_G  
    ;\]
    \item The final hybrid rigid transformation matrix is
    \[
    \pmb{Q} = \begin{bmatrix}
    \pmb{\mathcal{R}} & \pmb{\mathcal{T}} \\
    \mathbf{0}^T & 1
    \end{bmatrix}
    \]
\end{enumerate}

We also set $\det (\pmb{\mathcal{R}}) = 1$. \rev{For $\lambda \in [0,1]$ the interpolated quaternion $\mathbf{q}$ has unit norm, so $\pmb{\mathcal{R}}$ is an exact element of $SO(3)$. The constraint $\lambda \in [0,1]$ is enforced during training by the reparameterization described in Section~IV.}

\subsection{Network Design and Training}
\begin{figure*}[!bt]
\centering
 \includegraphics[width=1.0\textwidth]{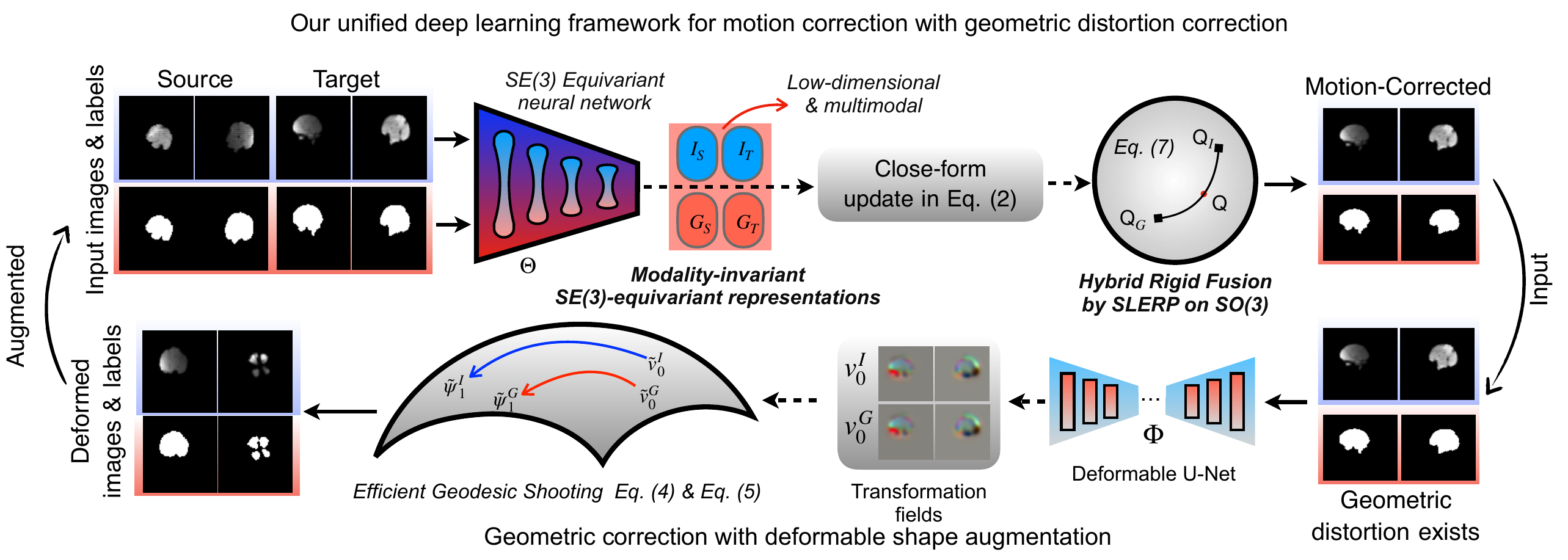}
     \caption{An illustration of the network architecture of our proposed motion correction learning framework, UniMo. \textbf{Top}: Motion correction is based on equivariant neural networks that take both images and segmentations as input. The low-dimensional equivariant spatial means of the images and segmentations are estimated simultaneously. An enhanced hybrid rigid transformation $Q$ is computed from both domains. The rigid loss function is calculated between the aligned and target images/labels. \textbf{Bottom}: A geometric shape augmenter (i.e., the deformation correction network described in Section~III-C) is incorporated into the U-Net based neural networks. It takes both aligned and target images/segmentations and estimates the transformation fields for both domains using separate loss functions for segmentations and images. } 
\label{fig: unimo}            
\end{figure*}

We develop a deep learning framework to estimate the objective expressed in Eq.~\eqref{eq:total}. Our framework comprises two major sub-modules: i) A hybrid rigid motion correction neural network, parameterized by equivariant filters, to produce $\pmb{Q}$; and ii) A hybrid deformation correction network, implemented using UNet, with deformable shape augmentation to estimate $\tilde{v}^I_0$ and $\tilde{v}^G_0$. Our framework is illustrated in Fig.~\ref{fig: unimo}. In the following sections, we provide a detailed description of the network architecture and the formulation of the network loss.
\paragraph*{\bf Hybrid Rigid Motion Correction Equivariant Network}
Let $\Theta = (\mathcal{T}_I, \mathcal{R}_I, \mathcal{T}_G, \mathcal{R}_G)$ represent the encoder parameters that learn rigid parameters from image and shape spaces, with $Q_I(\Theta)$ and $Q_G(\Theta)$ denoting the transformation functions yielded by the learned rigid parameters from the low-dimensional representations. Let $\pmb{Q}$ denote the hybrid rigid transformation. The rigid correction loss is computed between the aligned outputs and targets. Note that CNNs or recurrent neural networks can also be used for extracting low-dimensional features from images~\cite{evan2022keymorph, salehi2018real, singh2020deep}. However, such representations are limited in capturing the true nature of rigid transformations. In our approach, we employ an efficient method to compute rigid transformations within the equivariant neural network by calculating the equivariant spatial means of images~\cite{moyer2021equivariant}. We express the general formulation for the rigid motion network, which takes source and target images (or their segmentations), as follows:
\begin{align}
\label{eq:rigid_net} 
l(\Theta) = & \Vert S \circ \pmb{Q}(\Theta) - T \Vert_F, \quad \text{s.t.} \, \text{Eq.~\eqref{eq:closeform}} \,  \& \text{ Eq.~\eqref{eq: mmq}},
\end{align}
where \(\Vert\cdot\Vert_F\) denotes the Frobenius norm.

\paragraph*{\bf Deformation Correction Network (Geometric Shape Augmenter)}
We refer to the deformation correction network also as the \textit{geometric shape augmenter}, reflecting its dual role: (1) it estimates local deformation fields to correct for geometric distortions and non-rigid motion, improving overall alignment quality; and (2) the estimated deformation fields act as data augmentation during training, exposing the rigid correction network to a wider variety of spatial configurations.

Let \(\Phi\) denote the parameters of an encoder-decoder in our geometric learning network, where \(\psi_I(\Phi)\) represents the deformation fields and \(\tilde{v}_0(\Phi)\) represents the velocity fields learned by the network. The general formulation of the loss for the deformation estimation network, which takes aligned images (or segmentations) and target images (or segmentations), is given by:
\begin{align}
\label{eq:geo_net} 
l(\Phi) = & \frac{1}{\sigma^2} \Vert S \circ \pmb{Q}(\Theta) \circ \psi(\Phi) - T \Vert_2^2 + (\tilde{\mathcal{L}} \tilde{v}_0(\Phi), \tilde{v}_0(\Phi)) \nonumber \\
& + \text{reg}(\Theta, \Phi), \quad \text{s.t.} \, \text{Eq.~\eqref{eq:epdifffourier}} \, \& \text{ Eq.~\eqref{eq:lowtransformation}}.
\end{align}
Here, $\mathrm{reg}(\Theta, \Phi)$ is a regularization term on the network parameters, implemented as $L_2$ weight decay with a coefficient of $10^{-5}$.
In addition to estimating deformations using LDDMM~\cite{zhang2019fast}, we provide a deep learning model that learns stationary velocity fields~\cite{balakrishnan2019voxelmorph} while maintaining comparable model accuracy. Our framework is modular in the choice of the deformation backbone. While our default implementation uses LDDMM and VoxelMorph-based UNet, advanced predictive registration models such as TransMorph~\cite{chen2022transmorph} (a transformer-based method that uses self-attention for long-range spatial correspondence) and DiffuseMorph~\cite{kim2022diffusemorph} (a diffusion-model-based approach that generates deformation fields through iterative denoising) can also be integrated. These alternative backbones are not contributions of this work but illustrate the modularity of UniMo's design.
%Analogous to our previous work~\cite{wang2024spaer}, we extend UniMo to a spatio-temporal approach that incorporates time information to handle motion among data sequences.
\paragraph*{\bf Network Loss.}
The objective function for the hybrid model is defined as:
\begin{align}
\label{eq:total_net} 
l(\Phi, \Theta) = l_I(\Phi) + l_G(\Phi) + l_I(\Theta) + l_G(\Theta).
\end{align}
This total loss is the sum of four terms: $l_I(\Theta)$ and $l_G(\Theta)$ are the rigid correction losses computed from image and shape (geometric/segmentation) data respectively, while $l_I(\Phi)$ and $l_G(\Phi)$ are the deformation correction losses from image and shape data respectively. The subscripts $I$ and $G$ denote the Image and Geometric (shape) domains, and the parameters $\Theta$ and $\Phi$ correspond to the rigid and deformation networks.
We employ an alternating optimization scheme~\cite{nocedal1999numerical} to minimize the network loss. 
A summary of our joint learning process using alternating optimization is provided in Algorithm~\ref{alg1}.

\begin{algorithm}[h]
\SetKwSty{algokwsty}% arXiv: ieeeaccess.cls redefines \textbf, which breaks algorithm2e keywords
\DontPrintSemicolon
\SetAlgoLined
\SetArgSty{textnormal}
\SetKwInOut{Input}{Input}
\SetKwInOut{Output}{Output}

\Input{Source and target images with segmentation labels, number of iterations $h$.}
\Output{Motion parameters $\mathbf{Q}$, aligned image and segmentation, deformed image and segmentation, transformation fields.}

\Repeat{convergence}{
    \For{$i=1$ \textbf{to} $h$} {
        \tcc{Train Hybrid Motion Correction}
        Input the pair of images and segmentations into the equivariant network;\
        
        Compute the close-form update using Eq.~\eqref{eq:closeform} and Eq.~\eqref{eq: mmq} to produce the enhanced hybrid rigid transformation $\mathbf{Q}$\ by SLERP;\
        
        Output the aligned images and segmentations; minimize the motion correction loss defined in Eq.~\eqref{eq:rigid_net};\
        
        \tcc{Train Geometric Deformation Estimation}
        Input the aligned images and target images with segmentations into the registration network;\
        
        Output the velocity fields for both inputs and compute forward geodesic shooting using Eq.~\eqref{eq:epdifffourier} and Eq.~\eqref{eq:lowtransformation}, and backward shooting (Sec. B in supplementary);\
        
        Minimize the registration loss defined in Eq.~\eqref{eq:geo_net} and output the predicted velocity field and the deformed image.\
    }
}
\caption{Hybrid learning of UniMo.}
\label{alg1}
\end{algorithm}

\subsection{Extended Spatio-temporal Approach}
Following the approach in~\cite{wang2024spaer}, we extend UniMo to a spatio-temporal module for motion tracking in image time series. %, building on our prior work~\cite{wang2024spaer}. 
This module consists of 3 sub-modules: (i) a rigid motion correction network parameterized by equivariant filters, (ii) a temporal encoding module that incorporates time information of the sequence into the equivariant features, and (iii) a self-attention network that learns feature correspondence across different time points by combining features from (i) and (ii).

\paragraph*{\bf Spatio-temporal Closed-forms}
Given an image sequence $\pmb{I} = \{ I_0, I_1, ... , I_T \}$ changing over time $\pmb{t} =\{ 0, 1, ... , T \}$, we develop a spatio-temporal rigid transformation estimation to find the optimal path $\hat{\textbf{Q}}$ of rigid movement as follows, 
\begin{equation}
\label{eq: spaerenergy} 
E(\textbf{Q}) =  \sum^{T}_{t=0} \text{Dist} [I_t \circ Q_t(z_t, z_{t+1})  , I_{t+1}] ,
\end{equation} 
where $z_t$ represents the spatio-temporal representations for adjacent time frames. Here $\text{Dist} [\ \cdot \ ]$ denotes a distance term that measures the dissimilarity between the aligned and the target sequence. Analogous to ~\cite{moyer2021equivariant}, we derive the closed-form solution for both translation and rotation parameters between time point $t$ and $t+1$ to characterize $Q_t$,
\begin{equation}
\label{eq: spaercloseform} 
\mathcal{T}_t = z_{t+1} - \mathcal{R}_t z_{t},\quad \mathcal {R}_t = V_t \cdot {U_t}^T,\quad \text{s.t.} \, \, \det (\mathcal{R}_t) = 1,
\end{equation}
where $ {U}_t \Sigma_t V^*_t = z_{t} \cdot z_{t+1}^{T}$, $U_t$ and $V^*_t$ are real orthogonal matrices, $\Sigma_t$ is a diagonal matrix with non-negative real numbers on the diagonal. Setting the determinant of $\mathcal{R}_t$ to $1$ guarantees that it accurately reflects a rigid transformation. For further architectural details of the temporal encoding and self-attention components, we refer the reader to~\cite{wang2024spaer}.

\section{Experimental Evaluation}
\paragraph*{\bf Data}
For motion tracking in a single  domain (modality) test, we included $240$ sequences of 4D fetal echo-planar imaging (EPI) time series acquired on Siemens 3T scanners. The study was approved by the institutional review board and written informed consent was obtained from all participants. The gestational ages of the fetuses at the time of the scans were from $22.57$ to $38.14$ weeks (mean $32.39$ weeks). Imaging parameters included a slice thickness of $2$ to $3 mm$, a repetition time of $2$ to $5.6$s (mean $3.1$s), and an echo time of $0.03$ to $0.08$s (mean $0.04$s). Fetal brains were extracted from scans using a previously-validated segmentation method~\cite{faghihpirayesh2024fetal}, which provided segmentation maps used as shape input to UniMo. All brain scans were resampled to $96^3$ with a voxel size of $3 mm^3$ and underwent intensity normalization. 

For multiple, out-of-domain modality tests, in all baselines, we incorporated three different image datasets, including segmentation labels from varying organs, CT scans, and T1 MRIs, from publicly released medical image repositories. We used 60 CT scans from the Lung CT Segmentation Challenge (LCTSC)~\cite{yang2017data}. This dataset comprises 4DCT or free-breathing CT images (slice thickness of $2.5$ to $3$ mm) from 60 patients across three institutions, divided into 36 training datasets, 12 off-site test datasets, and 12 live test datasets. Manual segmentations are provided and serve as the ground truth, including the esophagus, heart, lungs, and spinal cord. For our work, we specifically extracted the left and right lungs for registration. Second, we included $200$ images ( slice thickness of 1 $mm^3$) from the MedMNIST datasets of varying organs~\cite{yang2021medmnist,medmnistv2}. The scans consisted of 3D CT scans of the adrenal gland, 3D CT scans of bone fractures, and 3D magnetic resonance angiography (MRA) scans of blood vessel shapes in the brain, as manually-segmented labels. We applied thresholding to pre-process all binary maps and converted them to distance transforms to obtain smooth shape representations; no spatial smoothing was applied to the images themselves. Third, for 3D  brain tumor MRI scans with tumor segmentation labels, we included $200$ public T1-weighted brain scans of different subjects from Brain Tumor Segmentation (BraTS) ~\cite{baid2023,baid2021rsna,menze2014multimodal} challenge 2021. All volumes from these datasets were resampled to $96^3$, with a voxel resolution of 1 $mm^3$, and underwent intensity normalization and bias field correction.

\paragraph*{\bf Motion Simulation}
Simulations for synthetic motion on real training and test data were performed on preprocessed, intensity-normalized image volumes. Rigid motion was modeled as six-degree-of-freedom transformations (translations in millimeters, rotations in degrees) applied about the image centroid. Non-rigid motion was modeled as B-spline free-form deformations on voxel grids with bounded control-point perturbations. Non-rigid fields were composed with rigid transforms, and images were resampled using linear interpolation.

Rigid motion was generated in two ways: 1) by interpolating a source--target transformation into $10$ steps and 2) by random-walk trajectories with Gaussian increments ($\sigma_T=0.5$--$2.0$ mm, $\sigma_R=0.2$--$1.0^{\circ}$) smoothed temporally by splines. Two regimens were used: small motion (translations $\leq 10$ mm, rotations $\leq 5^{\circ}$) and large motion (translations $\leq 30$ mm, rotations $\leq 20^{\circ}$). For fetal EPI, non-rigid fields were generated on B-spline grids with control point spacing of $8$--$12$ voxels (corresponding to $24$--$36$ mm at $3$ mm resolution) and perturbations $\leq 3$ mm. For lung CT, non-rigid fields were generated on grids with spacing of $10$ voxels ($10$ mm at $1$ mm resolution) and perturbations of $3$--$5$ mm, with diaphragm regions allowing up to $7$--$10$ mm to mimic respiratory deformation. For structural MRI, non-rigid fields were generated on grids with spacing of $10$ voxels ($10$ mm at $1$ mm resolution) and perturbations $\leq 3$ mm. Ground-truth frame-wise rigid transformations and non-rigid fields were recorded for quantitative evaluation of translational, angular, and overlap-based metrics.

\paragraph*{\bf Baselines \& Evaluation Metrics.}
We conducted three sets of experiments to evaluate UniMo: i) an ablation study of the weight parameter that balances the effect between shape and image, ii) motion correction performance on single image modality (fetal EPI), and iii) motion correction on the out-of-domain image modalities (LCTSC, MedMNIST, and BraTS) without network retraining.

First, we examined the estimated values of the weight parameter $\lambda$ with different initial values as convergence curves during network training. Next, we compared our model against state-of-the-art deep learning motion correction approaches, including DeepPose~\cite{salehi2018real}, KeyMorph~\cite{evan2022keymorph}, and Equivariant Filters~\cite{moyer2021equivariant}. Our evaluation includes both visual comparisons and quantitative analyses, focusing on translational and angular errors in data with simulated motions. Translational error is defined as the Euclidean distance (in mm) between the estimated and ground-truth translation vectors. Angular error is defined as the geodesic distance (in degrees) between the estimated and ground-truth rotation matrices~\cite{salehi2018real}. Additionally, we use a voxel-wise metric, temporal Signal to Noise Ratio (tSNR), to assess the quality and alignment of the EPI time series in the single modality task. tSNR is defined as $\mathrm{tSNR}(\mathbf{x}) = \mathrm{mean}(I(\mathbf{x},t)) / \mathrm{std}(I(\mathbf{x},t))$, where $I(\mathbf{x},t)$ is the voxel intensity at spatial location $\mathbf{x}$ across the time series. Higher tSNR values indicate greater temporal stability, reflecting better motion correction quality. Misalignment or motion degradation across an image time series lowers tSNR.

To test the sub-module of our spatial-temporal approach for motion tracking, we treated all baselines as static models to predict motion parameters between subsequent images of the time series. For motion correction and tracking, we tested all models on real scans with unknown motions, reporting the Dice coefficient between the target image and the aligned image. The Dice coefficient is computed between the segmentation of the motion-corrected source and the target segmentation. Specifically, after applying the estimated transformation to the source image, the same transformation is applied to the source segmentation, and the Dice overlap with the target segmentation is then computed. The segmentations used for Dice computation are: automatic segmentations from~\cite{faghihpirayesh2024fetal} for fetal brain data, manual lung segmentations provided by LCTSC, tumor labels from BraTS, and manually-segmented organ labels from MedMNIST. The effectiveness and stability of motion correction are demonstrated through a Dice coefficient analysis, comparing the alignment accuracy of images across different degrees of motion and sequence lengths.

\rev{\paragraph*{\bf Statistical Analysis}
Because all methods are evaluated on the same test pairs or sequences, differences between UniMo and each baseline were assessed with a two-sided paired Wilcoxon signed-rank test on the per-pair metric values. Multiple comparisons against the three baselines were controlled with the Holm-Bonferroni procedure, and $p < 0.05$ after correction was considered statistically significant. The test was applied to the translational error, angular error, and Dice coefficient in both the single-modality and cross-modality experiments.} %The $p$-values are reported in Table~\ref{tab:stats}. The spread of each metric across test pairs is shown by the distribution plots in Figs.~\ref{fig:stat}, \ref{fig:dice}, and \ref{fig:stat_MM}, which display the full per-pair distributions rather than point estimates.}

To evaluate the effectiveness of UniMo compared to baselines trained on multiple image modalities, we first report the translational and angular errors on simulated motions for all models. We present multiple examples, comparing all models across different image modalities. To demonstrate the advantage of UniMo, which does not require retraining with limited datasets, we also report mean Dice accuracy and the best epoch number for model training, including the deviation as the size of the training dataset increases. We note that %the Equivariant Filter baseline~\cite{moyer2021equivariant} is architecturally equivalent to UniMo with both the shape branch and the deformation augmenter removed (i.e., rigid correction using image intensities only). 
the difference in performance between the equivariant filter baseline~\cite{moyer2021equivariant} and UniMo serves as an evidence for the contribution of the shape integration and deformation augmentation components.

\paragraph*{\bf Implementation \& Parameters} 
For motion correction settings, we set the dimension of equivariant spatial means to $128$ when computing the close-form update of rigid transformation. We included a nine-layer equivariant neural network with two attention layers. For the geometric shape augmenter, we adopted a $7$-layer Unet and used $16$ as the reduced dimensionality of the low-dimensional frequencies in Eq.~\eqref{eq:epdifffourier}. We used $10$ time steps of Euler integration for geodesic shooting. \rev{The weight parameter $\lambda$ in Eq.~\eqref{eq: mmq} is learned during training. To keep $\lambda$ within $[0,1]$ while optimizing an unconstrained variable, we parameterize it as
\begin{equation}
\label{eq:lambda_param}
\lambda = \clip\!\left(\alpha / \lambda_{\max},\, 0,\, 1\right),
\end{equation}
where $\alpha$ is a trainable scalar updated by a separate Adam optimizer with a learning rate of $10^{-3}$, and $\lambda_{\max} = 20$ is a fixed scaling constant. The scaling by $\lambda_{\max}$ reduces the effective step size on $\lambda$ relative to that on $\alpha$, which prevents the fusion weight from changing abruptly early in training when the two rigid estimates are still inaccurate. All values of $\lambda$ reported in Section~V-A are the effective weight given by Eq.~\eqref{eq:lambda_param}. After training, $\lambda$ is fixed at its converged value $\lambda^{\ast}$ and is used as a constant at inference; it is not adapted per input pair, and the same $\lambda^{\ast}$ obtained from fetal EPI training was used for all unseen modalities in Section~V-C.} For network training, we used a batch size of $4$, a weight decay of $0.00001$ for $L_2$ regularization, and an initial learning rate of $\eta = 1 \times 10^{-5}$, with training conducted for $1000$ epochs using the Adam optimizer. The learning rate schedule employed cosine annealing to dynamically adjust the rate throughout training. The dataset was divided into $70\%$ for training, $15\%$ for validation, and $15\%$ for testing. The best-performing models were selected based on validation performance. All experiments were carried out on an NVIDIA RTX A6000 GPU with 48GB memory.

\section{Results}
\subsection{Ablation Study: Weight Parameter Analysis}
\begin{figure*}[!bt]
\centering
 \includegraphics[width=1.0\textwidth] {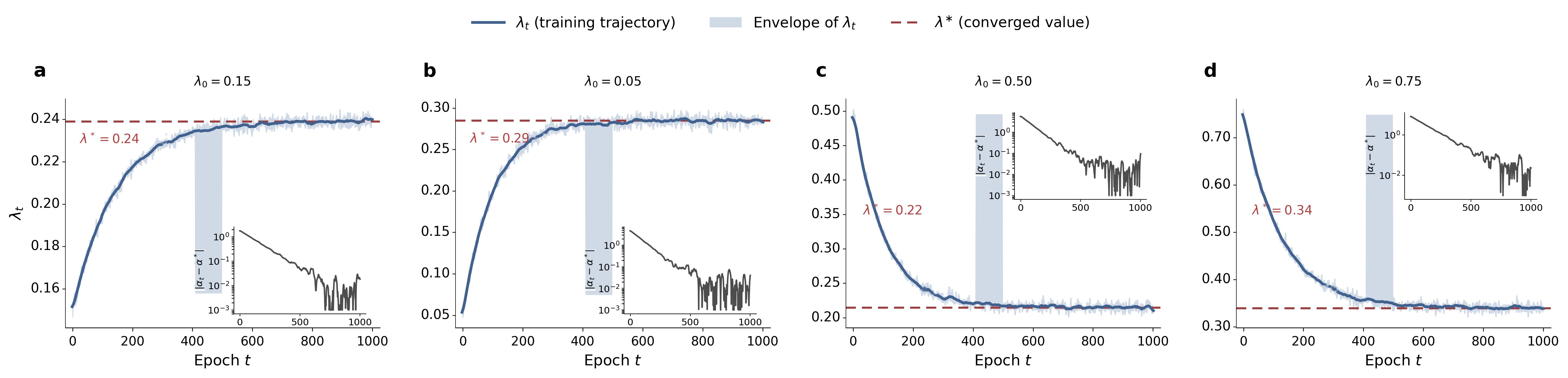}
 \caption{\rev{Convergence of the weight parameter $\lambda$ during training of
UniMo. \textbf{a--d},~Trajectories of the effective weight $\lambda_t$ defined by Eq.~\eqref{eq:lambda_param} over 1000 epochs of training
with the loss function~\eqref{eq:total_net}, under four initializations
$\lambda_0 = 0.15$, $0.05$, $0.50$, and $0.75$ (corresponding to $\alpha_0 = 3$, $1$, $10$, and $15$ with $\lambda_{\max}=20$), respectively. Solid curves show
the smoothed trajectory, shaded bands the per-epoch fluctuation envelope, and
dashed lines the final selected value $\lambda^{\ast}$. Insets: the deviation
$|\alpha_t - \alpha^{\ast}|$ of the unconstrained variable $\alpha = \lambda_{\max}\lambda$ on a logarithmic scale; the deviation of $\lambda_t$ is this quantity divided by $\lambda_{\max}$ and follows the same curve. The approximately
linear decay over the first ${\sim}500$ epochs indicates that $\lambda_t$
converges to $\lambda^{\ast}$ at an exponential rate, after which the residual
fluctuation is dominated by stochastic-gradient noise. Convergence is
observed regardless of the initialization, whether $\lambda_0$ is set below
or above $\lambda^{\ast}$.}}
\label{fig:convergence}            
\end{figure*}
Fig.~\ref{fig:convergence} shows the estimated $\lambda$ by UniMo with different initializations during model training. \rev{In all four cases $\lambda$ converged to a narrow range, $\lambda^{\ast} \in [0.22, 0.34]$ (dashed lines in Fig.~\ref{fig:convergence}), which lies strictly inside the admissible interval $[0,1]$ of Eq.~\eqref{eq: mmq}. Since $\lambda^{\ast} < 0.5$, the image branch receives the larger weight $1-\lambda^{\ast}$ in the fused rotation.} This consistent convergence indicates the robustness and reliability of the framework across various initial conditions. Importantly, $\lambda$ serves as the balancing weight within our hybrid SLERP-based rotation fusion strategy. \rev{The optimizer does not drive $\lambda$ to either endpoint from any initialization: it moves away from $0$ (image-only fusion) when initialized at $0.05$ and away from $1$ (shape-only fusion) when initialized at $0.75$. The converged value therefore corresponds to a fused rotation that lies at a point of the geodesic between the image-derived and shape-derived rotations, and indicates that the shape branch makes a consistent, nonzero contribution to the rigid estimate.}

\rev{Because $\lambda$ is learned by minimizing the training loss, a non-zero value of $\lambda^{\ast}$ means that adding the shape estimate reduces the loss below what the image branch achieves alone. The stability of $\lambda^{\ast}$ across initializations further supports the consistency of the SLERP fusion.} This makes UniMo a good choice for applications with significant intensity variations among time frames, such as prospective motion tracking in diffusion-weighted MRI, e.g.~\cite{verdera2025heron}.

\subsection{ Single Modality Evaluation}
\begin{figure*}[!bt]
\centering
 \includegraphics[width=1.0\textwidth] {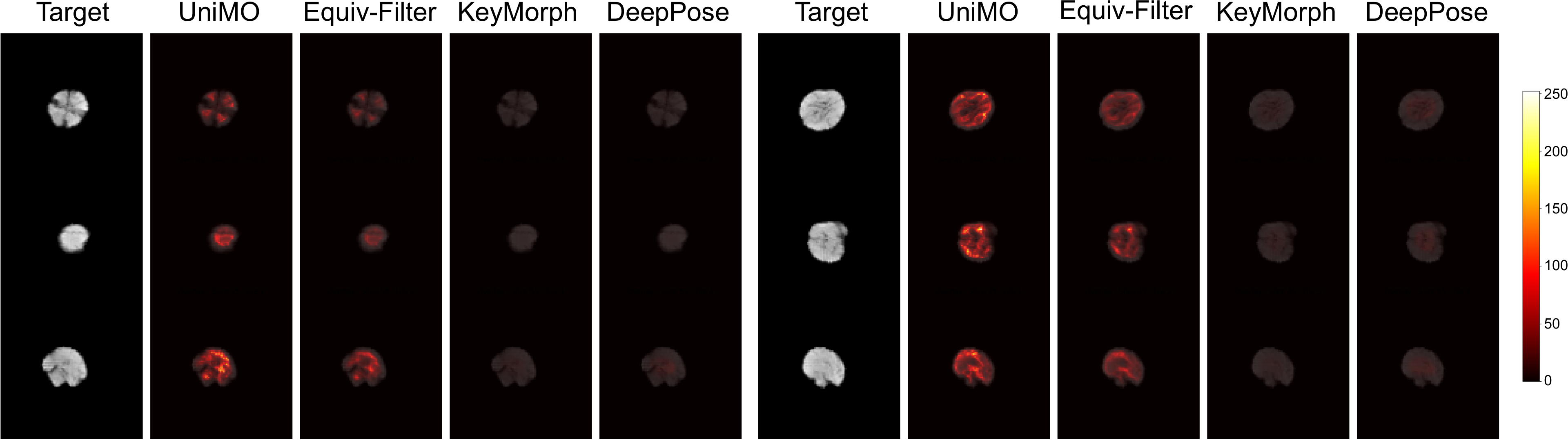}
 \caption{ Two examples of heat maps of tSNR estimated from all motion correction models over $50$ fetal EPI pairs. From left to right: Target image, heat maps from UniMo, Equiv-Filter, KeyMorph and DeepPose. Higher tSNR values indicate the best alignment of the image time series was obtained from UniMo.}
\label{fig:TSNR}            
\end{figure*}
Fig.~\ref{fig:TSNR} visualizes two cases of tSNR heat maps calculated over $50$ pairs for motion correction across all models. The heat maps demonstrate that UniMo displays higher tSNR values than other methods, indicating that UniMo accurately corrected motions and produced image time series with higher alignment accuracy than the alternative methods.

\begin{figure*}[!bt]
\centering
 \includegraphics[width=1.0\textwidth] {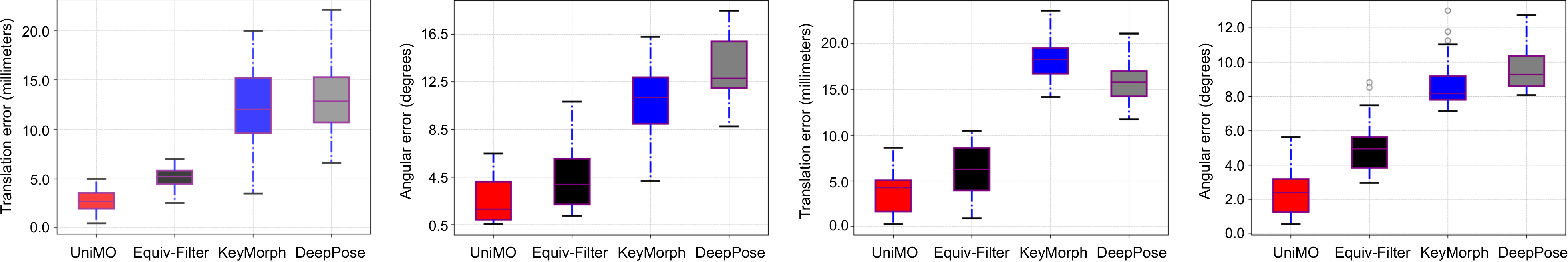}
 \caption{Statistical results for both \rev{translational} and angular errors of all models on single modality. Left: motion correction performance reported over $300$ image pairs; Right: motion tracking performance on 70 sequences of fetal EPI scans with simulated motions. \rev{Boxes show the median and interquartile range of the per-pair (left) and per-sequence (right) errors, whiskers extend to $1.5$ times the interquartile range, and circles mark outliers. $p$-values of the paired comparisons between UniMo and each baseline are given in Table~\ref{tab:stats}.}}
\label{fig:stat}            
\end{figure*}

The left side of Fig.~\ref{fig:stat} presents a comparative analysis of motion correction errors in translation and rotation across all methods. Our approach produced the lowest errors ($\sim$ $2.4$ mm of movement, and $\sim1.8^{\circ}$ of rotations for the fetal brain) with lowest variance between adjacent 3D volumes compared to other methods. \rev{The improvement over each baseline was statistically significant for both translational and angular error ($p < 0.05$ for all comparisons; Table~\ref{tab:stats}), including the next-best method, the equivariant filter model~\cite{moyer2021equivariant} ($p = 0.003$ for translation and $p = 0.011$ for rotation). It highlights the superior performance of UniMo in correcting fetal motions, consistently outperforming the other methods in accuracy.} 

The right side of Fig.~\ref{fig:stat} illustrates a comparison of motion tracking errors in translation and rotation across various methods. Our method achieves the lowest error rates for $70$ image sequences, with approximately $4.8$ mm for translational movements and $2.3$ degrees for rotational adjustments in fetal brain scans. UniMo exhibits lower alignment error between consecutive 3D volumes compared to the alternative methods, reflecting its superior motion tracking accuracy \rev{($p < 0.05$ against all baselines for both metrics; Table~\ref{tab:stats})}. Note that in this scenario on motion tracking on a 3D image time series, static models produced higher errors because they were not able to capture the long-term image sequence dependencies.

\begin{figure*}[!bt]
\centering
 \includegraphics[width=1.0\textwidth] {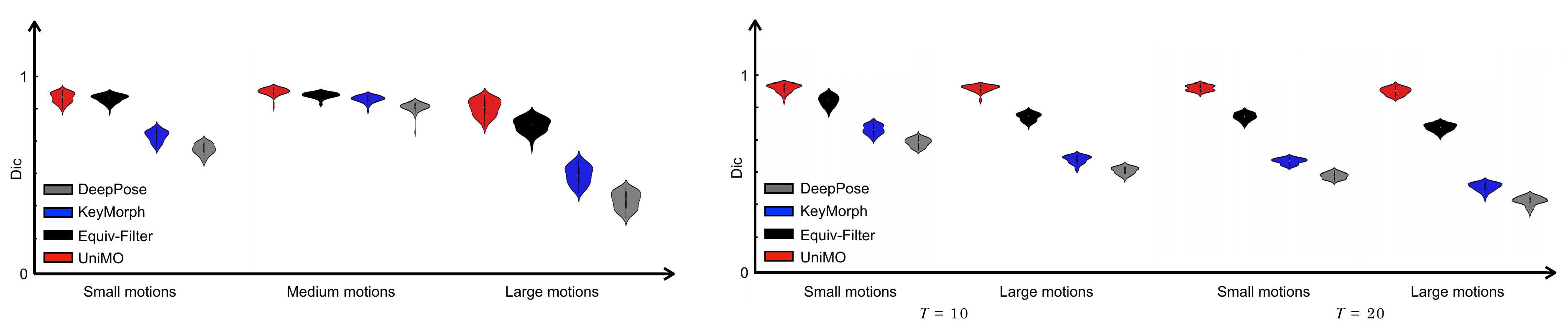}
 \caption{Statistical results of Dice comparison on fetal EPIs with unknown motions. Left: Motion correction performance with different degrees of motions, small ($\mathcal{T}_{max} = 10 \text{mm}$, $\mathcal{R}_{max} = 5^{\circ}$), medium ($\mathcal{T}_{max} = 20 \text{mm}, \mathcal{R}_{max} = 10^{\circ}$) and large motions ($\mathcal{T}_{max} = 30 \text{mm}$, $\mathcal{R}_{max} = 20^{\circ}$). The \rev{mean} dice score of our best model, for motion levels from left to right are, 0.97, 0.93, 0.92 \rev{(violins show the per-pair distribution of Dice values, with the inner marker at the median; $p$-values against each baseline in Table~\ref{tab:stats})}. Right: motion tracking performance across varying degrees and lengths of image sequences ($T$). Small ($\mathcal{T}_{max} = 10 \text{mm}$, $\mathcal{R}_{max} = 5^{\circ}$) and large motions ($\mathcal{T}_{max} = 30 \text{mm}$, $\mathcal{R}_{max} = 20^{\circ}$) were evaluated. Report efficiency with average time consumption: \textbf{0.501s} per pair / \textbf{9.960s} per sequence when $T=20$. }
\label{fig:dice}            
\end{figure*}

The right side of Fig.~\ref{fig:dice} quantitatively shows the accuracy of motion tracking comparison over varying degrees of motions and different lengths of data sequences. Our model exhibits superiority in handling real motions ranging from small to large, and it maintains comparable motion tracking accuracy when dealing with extended data sequences. This indicates the high stability and robustness of our model, as it demonstrates a high level of accuracy in correcting severe fetal brain movements. We also report the average time consumption for adjacent pairs and the entire sequence. The total computation time of our model for motion tracking is 10 seconds for a 4D sequence that takes $\sim60s$ to acquire. This paves the way for an efficient real-time tracking of the fetal head motion for prospective correction. 

The left side of Fig.~\ref{fig:dice} shows the dice coefficient comparisons for fetal EPI image pairs under various motion levels. It demonstrates that our method consistently achieves higher dice scores, regardless of the motion degree. \rev{Although the per-pair Dice distributions of UniMo and the equivariant filter baseline partially overlap, in particular at the small-motion level, the paired test shows that the difference is statistically significant at all three motion levels %($p = 0.033$, $0.043$, and $0.032$ for small, medium, and large motion against the equivariant filter model, and $p < 0.01$ against KeyMorph and DeepPose; 
($p<0.05$; Table~\ref{tab:stats})}. %The overlap in the violin plots reflects the range of Dice values across individual pairs, whereas the paired test evaluates the per-pair difference between methods on the same input.} 
\rev{The test statistics reported in Table~\ref{tab:stats} in combination with the summary statistics and per-pair distributions shown by the violin plots and the boxplots in Fig.~\ref{fig:stat} and Fig.~\ref{fig:dice} show that UniMo outperformed other methods. These results particularly highlight the robustness and stability of UniMo in challenging scenarios with significant motion.}

\begin{table}[!t]
\rev{\caption{$p$-values of the paired comparison between UniMo and Equiv-Filter~\cite{moyer2021equivariant}, KeyMorph~\cite{evan2022keymorph}, and DeepPose~\cite{salehi2018real} (two-sided paired Wilcoxon signed-rank test). Values below $0.001$ are reported as $<0.001$. All comparisons are significant at the $0.05$ level and remain so after Holm-Bonferroni correction over the three baselines within each row.}}
\centering
\footnotesize
\setlength{\tabcolsep}{4pt}
\begin{tabular}{lccc}
\toprule
Metric & Equiv-Filter & KeyMorph & DeepPose \\
\midrule
\multicolumn{4}{l}{\textit{Fetal EPI, motion correction ($n = 300$ pairs)}} \\
\quad Translational error & 0.003 & $<0.001$ & $<0.001$ \\
\quad Angular error & 0.011 & 0.010 & $<0.001$ \\
\addlinespace
\multicolumn{4}{l}{\textit{Fetal EPI, motion tracking ($n = 70$ sequences)}} \\
\quad Translational error & 0.001 & 0.002 & $<0.001$ \\
\quad Angular error & 0.023 & 0.001 & $<0.001$ \\
\addlinespace
\multicolumn{4}{l}{\textit{Fetal EPI, real motion, Dice}} \\
\quad Small motion & 0.033 & 0.004 & $<0.001$ \\
\quad Medium motion & 0.043 & 0.008 & $<0.001$ \\
\quad Large motion & 0.032 & 0.001 & $<0.001$ \\
\addlinespace
\multicolumn{4}{l}{\textit{Cross-modality, motion correction}} \\
\quad Translational error & 0.003 & $<0.001$ & $<0.001$ \\
\quad Angular error & 0.002 & $<0.001$ & $<0.001$ \\
\bottomrule
\end{tabular}
\label{tab:stats}
\end{table}

\subsection{ Cross-modality Evaluation}

\begin{figure*}[!bt]
\centering
 \includegraphics[width=1\textwidth] {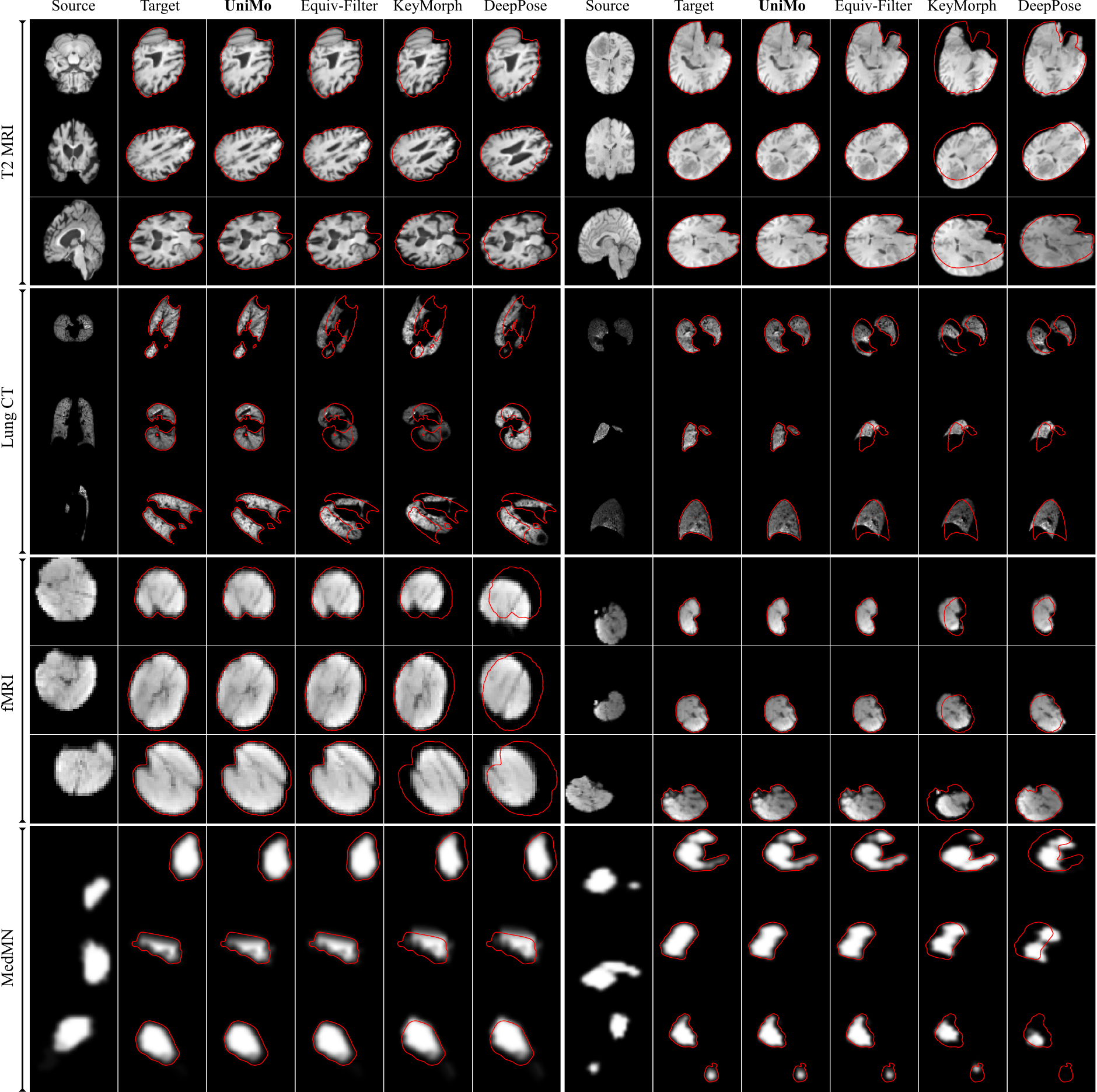}
 \caption{Motion correction comparison across multiple image modalities for all models. The image modalities from top to bottom are: T1 MRI scans of brains containing lesions, lung CT scans, EPI scans of fetal brains, and the shapes of adrenal glands. The images from left to right are: source, target, and aligned images by UniMo, Equivariant Filter~\cite{moyer2021equivariant}, KeyMorph~\cite{evan2022keymorph}, and DeepPose~\cite{salehi2018real}. The aligned images are displayed with red contours for better visualization.}
\label{fig:mm}            
\end{figure*}

Fig~\ref{fig:mm} shows eight cases of motion correction comparisons across multiple modalities, including T1 MRI, fetal EPI scans, lung CT scans, and organ shapes from MedMNIST. Our method, trained on a single modality, consistently outperforms all methods trained on multiple modalities. Specifically, for T1 MRI, fetal EPI scans, and organ shapes from MedMNIST, our method significantly outperforms KeyMorph and DeepPose and slightly surpassed the equivariant filter model. For lung CT scans, where image intensities exhibited significant contrast variations, UniMo demonstrated superior performance compared to all other baselines. This is because other models tend to neglect shape information when estimating motion parameters, focusing predominantly on image intensities. In contrast, UniMo effectively integrates shape information, leading to more accurate results. The visualization demonstrates that our model can be effectively employed across various modalities with high accuracy.

\begin{figure*}[!bt]
\centering
 \includegraphics[width=1.0\textwidth] {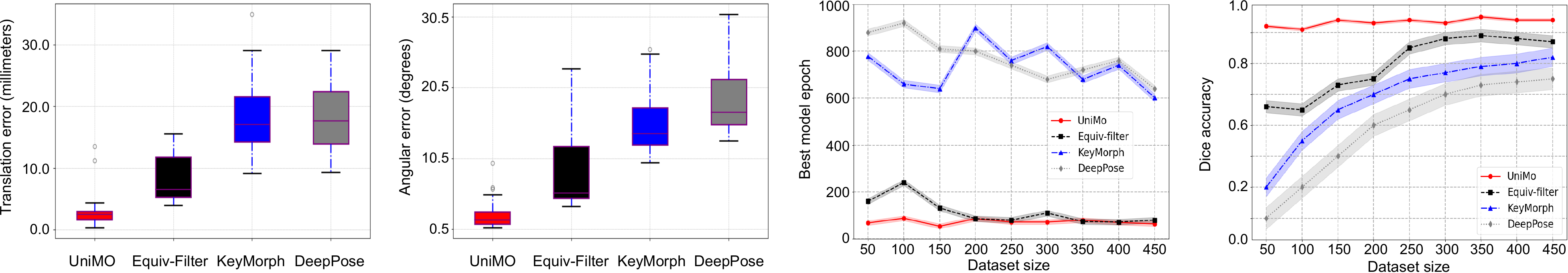}
 \caption{From left to right: \rev{translational} and angular errors of motion correction of four image modalities, epoch number of best model performance from training with varying training dataset size, and quantitative report of Dice accuracy with varying training dataset size. \rev{In the two left panels, boxes show the median and interquartile range of the per-pair errors, whiskers extend to $1.5$ times the interquartile range, and circles mark outliers.} }
\label{fig:stat_MM}            
\end{figure*}

The left side of Fig.~\ref{fig:stat_MM} presents the statistical comparison of motion correction across all models. UniMo demonstrates the lowest error, with an average error of $2.51$ mm and a rotation of about $1.9$ degrees, across various image modalities. \rev{The differences with respect to all three baselines were statistically significant for both metrics ($p < 0.005$; Table~\ref{tab:stats}). The margin over the equivariant filter model was smallest on T1 MRI and MedMNIST, where intensity contrast between source and target is consistent, and largest on lung CT, where it is not.}
These results partially highlight the benefits of incorporating shape information that allowed UniMo achieve high performance across different image modalities, whereas the other techniques performed poorly.
The right side of Fig.~\ref{fig:stat_MM}  illustrates the optimal epoch of model training and Dice accuracy for various training dataset sizes. Our model demonstrates rapid convergence and maintains consistently high accuracy even with datasets smaller than $50$ samples. In contrast, baseline models such as KeyMorph and DeepPose show gradual improvements in accuracy as the training data size increases. This is attributed to their limited capacity to accurately learn and generalize the nature of rigid motion across different image modalities. Although the original equivariant filter model performs relatively well when the dataset size is reduced to one third, its average accuracy is still compromised in cases where image intensities exhibit significant variations. UniMo effectively captures the inherent characteristics of rigid transformations and remains robust against fluctuations in image intensities and contrast. This highlights the superior generalizability of UniMo across diverse imaging conditions.

\section{Discussion}
Image registration remains the cornerstone of both retrospective and prospective motion correction in medical imaging. Recent breakthroughs in deep learning have redefined capabilities in this field. UniMo advances the field by integrating shape and intensity information and leveraging advanced neural network architectures to achieve a robust solution to motion correction. It addresses persistent limitations in existing frameworks, offering a reliable solution with low-latency inference that is required for real-time motion monitoring and prospective correction. 

The clinical utility of this approach is most evident in challenging applications such as fetal MRI, where large, non-periodic motion frequently compromises the quality of images~\cite{calixto2024advances,afacan2019fetal,christiaens2019utero}. Beyond improving clinical measurement of anatomical structures, e.g.~\cite{pier20163d}, motion correction is a prerequisite for advanced imaging, including fetal cardiac MRI~\cite{roy2019fetal,van2019fetal,vollbrecht2024fetal,vollbrecht2025improving}, 4D flow MRI~\cite{roberts2020fetal,van2023fetal,tompkins2025third},  diffusion-weighted MRI~\cite{marami2017temporal,snoussi2025haitch,verdera2025heron}, and functional MRI ~\cite{sobotka2022motion,neves2023real,taymourtash2025measuring}. By mitigating the motion barriers that have historically sidelined these techniques in critical populations, UniMo can serve as an enabler for more sophisticated diagnostic and prognostic assessments. Future efforts will focus on embedding this framework into prospective motion tracking and retrospective image processing pipelines.

The low-latency and rapid inference provided by deep learning models, such as UniMo, enable real-time motion monitoring and prospective navigation. Integrating motion tracking into the acquisition workflow ensures that data maintains sufficient coverage and quality for robust image reconstruction. In functional MRI, real-time monitoring is used to extend time-series acquisitions, compensating for motion-corrupted samples that must be censored during analysis~\cite{thesen2000prospective,dosenbach2017real,sui2020slimm,neves2023real}. Similarly, in structural and diffusion-weighted MRI, real-time motion detection or monitoring allows for dynamic correction of the acquisition plane or targeted reacquisition of corrupted data~\cite{benner2011diffusion,slipsager2021comparison,gagoski2022automated,verdera2025heron}. UniMo is particularly well-suited for these applications because it estimates bulk rigid motion while simultaneously accounting for deformations and geometric distortions. 

\rev{The inference time of UniMo is $0.5$ s per image pair. For volumetric EPI, motion is estimated once per acquired volume, so the relevant reference is the repetition time, which was $2$ to $5.6$ s (mean $3.1$ s) in our fetal data and is typically $1$ to $3$ s in adult functional MRI. An estimate that is available within $0.5$ s of volume reconstruction therefore leaves most of one repetition period for updating the slice prescription before the next volume is excited, which is the operating regime of volume-to-volume prospective correction~\cite{thesen2000prospective,sui2020slimm}. For navigator-based structural and diffusion acquisitions, in which a navigator volume is acquired every few seconds and the correction is applied to the subsequent readout, the same latency is also sufficient. UniMo is not fast enough for slice-to-slice correction, in which the update interval is on the order of tens to a few hundred milliseconds; this would require further acceleration, for example by reducing the input resolution or by pruning the equivariant network.}

\rev{For a comparison of deep learning models against classical optimization-based registration methods, we refer to \cite{billot2024se}. In this work, an equivariant CNN was compared in terms of accuracy and speed of inference (latency) against intensity-based rigid registration using ANTS~\cite{avants2011reproducible} and robust point matching (RPM)~\cite{jian2010robust}. Overall, the results show that the accuracy of the classical methods was comparable to the best deep learning model (equivariant CNN), however, classical registration methods showed 1) higher latency than the deep learning methods (4.2-7.3s compared to 0.5s), and 2) variable performance across modalities. A careful comparison of these two approaches for rigid and non-rigid registration requires implementation and optimization of both types of the classical methods on GPUs, so that the methods can be compared on the same hardware. We have done some work on GPU-based implementation and optimization of classical registration methods, however, additional work is needed 1) for a comparable implementation of classical, iterative deformable registration on GPU and 2) to conduct a fair and careful comparison of classical methods against deep learning methods on multi-modal medical imaging data. It is noteworthy that, in general, classical deformable registration is much slower than both classical rigid registration and (rigid and deformable) deep learning-based methods.}

\rev{UniMo is designed purposely to handle a broad spectrum of motion types commonly encountered in clinical imaging, including rigid transformations (e.g., head rotations~\cite{kober2011head,wallace2024rapid}, patient repositioning~\cite{wasza2012real}), periodic non-rigid motion~\cite{qi2020non} (e.g., respiratory and cardiac motion), and gradual anatomical changes~\cite{rong2021rigid} (e.g., tumor growth, brain atrophy, or edema).} It can be used with a wide range of imaging techniques, most notably non-Cartesian sampling and advanced echo-planar imaging techniques. While these techniques possess inherent motion robustness, UniMo may provide a critical secondary layer of correction for residual motion and complex geometric distortions. By leveraging shape information, UniMo maintains registration accuracy even when intensity-based features are inconsistent or degraded. This enables more reliable alignment for longitudinal studies~\cite{lv2022joint} (e.g., monitoring neurodegeneration or treatment response), correction of motion and geometric distortion artifacts in dynamic imaging~\cite{kurugol2017motion,timms2026fast} (e.g., abdominal or thoracic scans), and improved fusion across sessions or modalities. Robust performance for various motion types can support a range of applications spanning from longitudinal analysis to diagnosis, and real-time interventions.

\rev{\subsection{Limitations}
Several aspects of this work should be kept in mind when interpreting the results. The quantitative evaluation relies on simulated motion to provide a reference transformation. Although the fetal EPI test pairs are taken from different time frames of the same subject and therefore contain real motion, the accuracy of the method on real motion has not been measured against an independent reference, and effects that arise during acquisition rather than in the reconstructed volume are not represented in the simulation. While the increase in both tNSR and Dice metrics indicate improved alignment of the image time series, the results should still be interpreted as evidence of registration accuracy rather than measures of clinical performance.

The method also requires a segmentation of the structure of interest for both training and inference. This is available in many settings through automatic segmentation, but the quality of the shape branch is tied to the quality of that segmentation, and the sensitivity of the method to segmentation errors was not studied here. Finally, the comparison is limited to learning-based methods, and the fusion weight was learned on a single modality and applied unchanged to the others. A careful comparison with classical registration tools, optimized and implemented to run on the same hardware, and evaluation in clinical settings are natural next steps.}

\section{Conclusion}

In this paper, we introduced UniMo, a Unified Motion Correction framework as a framework that leverages deep neural networks to address the challenges of motion correction across various imaging datasets. By employing advanced neural network architectures with equivariant filters, UniMo overcomes the limitations of existing methods that require iterative retraining for each new image modality. Instead, UniMo achieves remarkable stability and adaptability through a single training phase on one modality, enabling effective application across multiple unseen modalities. UniMo excels by integrating hybrid knowledge from both shape and image data, significantly improving motion correction accuracy despite variations in image appearance. The inclusion of a geometric deformation augmenter enhances robustness by mitigating local geometric distortions and generating augmented data, thereby refining the training process. Experimental results across four distinct image datasets with various modalities demonstrate that UniMo outperforms current motion correction methods, representing a substantial advancement in medical imaging, particularly for complex image-guided motion correction applications such as real-time fetal head motion tracking. Looking ahead, several promising avenues for future work emerge. Firstly, exploring real-time applications and integrating UniMo with other imaging technologies, such as image segmentation, could enhance its utility in dynamic environments. Secondly, investigating advanced methods for selecting and utilizing different shape descriptors within UniMo may further boost joint learning performance and overall effectiveness. \rev{Thirdly, evaluation on data with real motion across the non-fetal modalities, comparison with classical registration tools, and a systematic study of sensitivity to segmentation quality will be needed to establish the clinical operating range of the method.}

\section{Appendix A }
%In this supplementary document, we provide additional mathematical derivations and implementation details that complement the main manuscript. Section~\ref{app: matrix} details the conversions and interpolations regarding motion correction quaternions and rotation matrices. Section~\ref{app: deformation} outlines the gradient derivations for the forward-backward geodesic shooting approach in deformation estimation.
\subsection{Derivations of Motion Correction}
\label{app: matrix}
\begin{itemize}
    \item Conversion from a rotation matrix to quaternion:\\

    Given a rotation matrix \( \mathcal{R} = \begin{bmatrix} R_{00} & R_{01} & R_{02} \\ R_{10} & R_{11} & R_{12} \\ R_{20} & R_{21} & R_{22} \end{bmatrix} \), the corresponding quaternion \( \mathbf{q} = [x, y, z, w] \) can be computed as follows:

\[
\begin{aligned}
    w &= \sqrt{1 + R_{00} + R_{11} + R_{22}} / 2, \\
    x &= \frac{R_{21} - R_{12}}{4w}, \\
    y &= \frac{R_{02} - R_{20}}{4w}, \\
    z &= \frac{R_{10} - R_{01}}{4w}.
\end{aligned}
\]
\item Spherical linear interpolation:\\

Given two quaternions \( \mathbf{q}_I = [x_I, y_I, z_I, w_I] \) and \( \mathbf{q}_G = [x_G, y_G, z_G, w_G] \) derived from image and shape domains, and a weight \( \lambda \rev{\in [0,1]} \). We derive operations,
\[
<\mathbf{q}_I, \mathbf{q}_G> = x_I x_G + y_I y_G + z_I z_G + w_I w_G,
\]
\[
(\mathbf{q}_I, \mathbf{q}_G)^{\Lambda} = \max(\min(\text{dot}(\mathbf{q}_I, \mathbf{q}_G), 1.0), -1.0)
\]
where $<\cdot,\cdot>$ denotes the dot product and $(\cdot,\cdot)^{\Lambda}$ denotes clipping the dot product to ensure it lies within the valid range. 

We calculate the angle \( \theta \) between the quaternions and the interpolation:
\[
\theta = \text{acos}((\mathbf{q}_I, \mathbf{q}_G)^{\Lambda} )
\]
\[
\mathbf{q} = \frac{\sin((1 - \lambda) \theta)}{\sin(\theta)} \mathbf{q}_I + \frac{\sin(\lambda \theta)}{\sin(\theta)} \mathbf{q}_G
\]
\rev{For $\lambda \in [0,1]$ the result is a unit quaternion on the shorter great-circle arc between $\mathbf{q}_I$ and $\mathbf{q}_G$. If the dot product is negative, $\mathbf{q}_G$ is replaced by $-\mathbf{q}_G$ before interpolation, which represents the same rotation and ensures the shorter arc is taken. When $\theta$ is close to zero, the formula is replaced by normalized linear interpolation to avoid division by a small $\sin(\theta)$.}

\item Conversion from Quaternion to Rotation Matrix:\\

Given an interpolated \rev{multimodal} quaternion \( \mathbf{q} = [x, y, z, w] \), convert it back to a rotation matrix:
\[
\pmb{\mathcal{R}} = \begin{bmatrix}
1 - 2y^2 - 2z^2 & 2xy - 2wz & 2xz + 2wy \\
2xy + 2wz & 1 - 2x^2 - 2z^2 & 2yz - 2wx \\
2xz - 2wy & 2yz + 2wx & 1 - 2x^2 - 2y^2
\end{bmatrix}
\]
\end{itemize}
\subsection{Derivations of Deformation Estimation}
\label{app: deformation}
Note that the deformation correction loss comprises contributions from both the image and segmentation estimations. In the following section, we focus on the derivation for $\tilde{v}_I$ as an example; the same process applies to $\tilde{v}_G$. While we explore three different methods~\cite{hinkle2018lagomorph,balakrishnan2019voxelmorph,zhang2019fast} for generating deformations within this model, we provide detailed derivations for the forward-backward shooting approach.

We initialize the geodesic shooting with the network output $\tilde{v}_0^G(\theta)$ and utilize a forward-backward shooting technique~\cite{wang2023metamorph,zhang2019fast} that incorporates adjoint Jacobi fields in Fourier space. The gradient of the loss function with respect to the predicted initial velocity fields $\tilde{v}_{G}$ before back-propagation is derived as follows:
\begin{enumerate}
\item 
Forward integrating the geodesic shooting equation (a.k.a. EPDiff) in Eq.(2) to compute $\tilde{v}_{G}(\rm{\Theta})_{t=1}$ at time point $t=1$ after obtaining the predicted initial velocity fields  $\tilde{v}_0^{G}$ from network forward-propagation;  
\item Compute the gradient of the loss function $l_(\mathrm{\Theta})$ with respect to $\tilde{v}_{G}(\rm{\Theta})_{t=1}$, 
\begin{align}
    &\nabla_{\tilde{v}_{G}(\rm{\Theta})_{t=1}} l(\Theta) = \nonumber \\ &\tilde{\mathcal{K}} \left( \frac{1}{\sigma^2_j}(S \circ \psi_1 [ \tilde{v}_{G}(\rm{\Theta})_{t=1}]  - T) \cdot \nabla (S \circ \psi [ \tilde{v}_{G}(\rm{\Theta})_{t=1}] \right); \nonumber
    \label{eq:gradvq1}
\end{align}
\item Bring the gradient in (ii) back to the space of initial velocity fields defined at the time point $t=0$ by integrating adjoint Jacobi fields backward in time obtain $\nabla_{\tilde{v}_{nj}(\rm{\Theta}_g)} l_{\text{Geo}}$,
\begin{equation} 
    \frac{d{\hat{v}}}{dt} = -\ad ^{\dagger}_{\tilde{v}}{\hat{u}}, \quad
    \frac{d{\hat{u}}}{dt} = -\hat{v} -\ad_{\tilde{v}} \hat{u} + \ad ^{\dagger}_{\hat{u}} \tilde{v}, \nonumber
    \label{eq:adjacobi}
    \end{equation}
    where $\hat{v} \in V$ are introduced adjoint variables with an initial condition $\hat{u} = 0, \hat{v} =  \nabla_{\tilde{v}_n(\rm{\Theta}_g)_1} l_{\text{Geo}}$ at $t=1$.  Here $\ad ^{\dagger}$ is an adjoint operator to the negative Lie bracket of vector fields, $\ad_{\tilde{v}} \tilde{w} = -[\tilde{v}, \tilde{w}]
= \tilde{\mathcal{D}}\tilde{v} \ast \tilde{w} - \tilde{\mathcal{D}}\tilde{w} \ast \tilde{v}$.
\item We backward propagate the gradient $\nabla_{\tilde{v}_{nj}(\rm{\Theta}_g)} l_{\text{Geo} }$ to geometric shape learning network. 
\end{enumerate}

\bibliographystyle{IEEEtran}
\bibliography{UniMo2025}

\begin{thebibliography}{100}
\providecommand{\url}[1]{#1}
\csname url@samestyle\endcsname
\providecommand{\newblock}{\relax}
\providecommand{\bibinfo}[2]{#2}
\providecommand{\BIBentrySTDinterwordspacing}{\spaceskip=0pt\relax}
\providecommand{\BIBentryALTinterwordstretchfactor}{4}
\providecommand{\BIBentryALTinterwordspacing}{\spaceskip=\fontdimen2\font plus
\BIBentryALTinterwordstretchfactor\fontdimen3\font minus
  \fontdimen4\font\relax}
\providecommand{\BIBforeignlanguage}[2]{{%
\expandafter\ifx\csname l@#1\endcsname\relax
\typeout{** WARNING: IEEEtran.bst: No hyphenation pattern has been}%
\typeout{** loaded for the language `#1'. Using the pattern for}%
\typeout{** the default language instead.}%
\else
\language=\csname l@#1\endcsname
\fi
#2}}
\providecommand{\BIBdecl}{\relax}
\BIBdecl

\bibitem{zaitsev2015motion}
M.~Zaitsev, J.~Maclaren, and M.~Herbst, ``Motion artifacts in mri: A complex
  problem with many partial solutions,'' \emph{Journal of Magnetic Resonance
  Imaging}, vol.~42, no.~4, pp. 887--901, 2015.

\bibitem{maclaren2013prospective}
J.~Maclaren, M.~Herbst, O.~Speck, and M.~Zaitsev, ``Prospective motion
  correction in brain imaging: a review,'' \emph{Magnetic resonance in
  medicine}, vol.~69, no.~3, pp. 621--636, 2013.

\bibitem{kyme2021motion}
A.~Z. Kyme and R.~R. Fulton, ``Motion estimation and correction in spect, pet
  and ct,'' \emph{Physics in Medicine \& Biology}, vol.~66, no.~18, p. 18TR02,
  2021.

\bibitem{silva2018challenges}
M.~A. Silva, A.~P. See, W.~I. Essayed, A.~J. Golby, and Y.~Tie, ``Challenges
  and techniques for presurgical brain mapping with functional mri,''
  \emph{NeuroImage: Clinical}, vol.~17, pp. 794--803, 2018.

\bibitem{runge2019motion}
V.~M. Runge, J.~K. Richter, and J.~T. Heverhagen, ``Motion in magnetic
  resonance: new paradigms for improved clinical diagnosis,''
  \emph{Investigative radiology}, vol.~54, no.~7, pp. 383--395, 2019.

\bibitem{wallace2021free}
T.~E. Wallace, O.~Afacan, C.~Jaimes, J.~Rispoli, K.~Pelkola, M.~Dugan,
  T.~Kober, and S.~K. Warfield, ``Free induction decay navigator motion metrics
  for prediction of diagnostic image quality in pediatric mri,'' \emph{Magnetic
  resonance in medicine}, vol.~85, no.~6, pp. 3169--3181, 2021.

\bibitem{malamateniou2013motion}
C.~Malamateniou, S.~Malik, S.~Counsell, J.~Allsop, A.~McGuinness, T.~Hayat,
  K.~Broadhouse, R.~Nunes, A.~Ederies, J.~Hajnal \emph{et~al.},
  ``Motion-compensation techniques in neonatal and fetal mr imaging,''
  \emph{American Journal of Neuroradiology}, vol.~34, no.~6, pp. 1124--1136,
  2013.

\bibitem{jaimes2016strategies}
C.~Jaimes and M.~S. Gee, ``Strategies to minimize sedation in pediatric body
  magnetic resonance imaging,'' \emph{Pediatric radiology}, vol.~46, no.~6, pp.
  916--927, 2016.

\bibitem{harrington2022strategies}
S.~G. Harrington, C.~Jaimes, K.~M. Weagle, M.-L.~C. Greer, and M.~S. Gee,
  ``Strategies to perform magnetic resonance imaging in infants and young
  children without sedation,'' \emph{Pediatric radiology}, vol.~52, no.~2, pp.
  374--381, 2022.

\bibitem{uus2023retrospective}
A.~U. Uus, A.~Egloff~Collado, T.~A. Roberts, J.~V. Hajnal, M.~A. Rutherford,
  and M.~Deprez, ``Retrospective motion correction in foetal mri for clinical
  applications: existing methods, applications and integration into clinical
  practice,'' \emph{The British journal of radiology}, vol.~96, no. 1147, p.
  20220071, 2023.

\bibitem{wang2019echo}
F.~Wang, Z.~Dong, T.~G. Reese, B.~Bilgic, M.~Katherine~Manhard, J.~Chen, J.~R.
  Polimeni, L.~L. Wald, and K.~Setsompop, ``Echo planar time-resolved imaging
  (epti),'' \emph{Magnetic resonance in medicine}, vol.~81, no.~6, pp.
  3599--3615, 2019.

\bibitem{dong2025romer}
Z.~Dong, T.~G. Reese, H.-H. Lee, S.~Y. Huang, J.~R. Polimeni, L.~L. Wald, and
  F.~Wang, ``Romer-epti: rotating-view motion-robust super-resolution epti for
  snr-efficient distortion-free in-vivo mesoscale diffusion mri and
  microstructure imaging,'' \emph{Magnetic resonance in medicine}, vol.~93,
  no.~4, pp. 1535--1555, 2025.

\bibitem{dai2025instantaneous}
C.~Dai, C.~Cai, J.~Wu, L.~Zhu, X.~Qu, Q.~Yang, J.~Zhou, and S.~Cai,
  ``Instantaneous t 2 mapping via reduced field of view multiple
  overlapping-echo detachment imaging: Application in free-breathing abdominal
  and myocardial imaging,'' \emph{IEEE Transactions on Biomedical Engineering},
  2025.

\bibitem{yang2026rapid}
Q.~Yang, J.~Bao, Z.~Ni, J.~Wang, L.~Chen, S.~Yu, Z.~Chen, C.~Cai, and S.~Cai,
  ``Rapid flow-artifact-free high-resolution t2 mapping via multi-shot multiple
  overlapping-echo detachment imaging,'' \emph{Magnetic Resonance in Medicine},
  vol.~95, no.~3, pp. 1393--1409, 2026.

\bibitem{ge2026high}
N.~Ge, Q.~Yang, C.~Yan, F.~Wu, Y.~Zhang, D.~Wu, Z.~Chen, J.~Zhong, C.~Cai,
  J.~Bao \emph{et~al.}, ``High-fidelity t2 relaxometry of the developing fetal
  brain using an 8-second single-shot multiple overlapping-echo detachment
  (moled) mri technique,'' \emph{Journal of Magnetic Resonance Imaging},
  vol.~63, no.~3, pp. 748--758, 2026.

\bibitem{deshmane2012parallel}
A.~Deshmane, V.~Gulani, M.~A. Griswold, and N.~Seiberlich, ``Parallel mr
  imaging,'' \emph{Journal of Magnetic Resonance Imaging}, vol.~36, no.~1, pp.
  55--72, 2012.

\bibitem{lustig2007sparse}
M.~Lustig, D.~Donoho, and J.~M. Pauly, ``Sparse mri: The application of
  compressed sensing for rapid mr imaging,'' \emph{Magnetic Resonance in
  Medicine: An Official Journal of the International Society for Magnetic
  Resonance in Medicine}, vol.~58, no.~6, pp. 1182--1195, 2007.

\bibitem{pipe1999motion}
J.~G. Pipe, ``Motion correction with propeller mri: application to head motion
  and free-breathing cardiac imaging,'' \emph{Magnetic Resonance in Medicine:
  An Official Journal of the International Society for Magnetic Resonance in
  Medicine}, vol.~42, no.~5, pp. 963--969, 1999.

\bibitem{jung2009k}
H.~Jung, K.~Sung, K.~S. Nayak, E.~Y. Kim, and J.~C. Ye, ``k-t focuss: a general
  compressed sensing framework for high resolution dynamic mri,''
  \emph{Magnetic Resonance in Medicine: An Official Journal of the
  International Society for Magnetic Resonance in Medicine}, vol.~61, no.~1,
  pp. 103--116, 2009.

\bibitem{usman2013motion}
M.~Usman, D.~Atkinson, F.~Odille, C.~Kolbitsch, G.~Vaillant, T.~Schaeffter,
  P.~G. Batchelor, and C.~Prieto, ``Motion corrected compressed sensing for
  free-breathing dynamic cardiac mri,'' \emph{Magnetic resonance in medicine},
  vol.~70, no.~2, pp. 504--516, 2013.

\bibitem{feng2014golden}
L.~Feng, R.~Grimm, K.~T. Block, H.~Chandarana, S.~Kim, J.~Xu, L.~Axel, D.~K.
  Sodickson, and R.~Otazo, ``Golden-angle radial sparse parallel mri:
  combination of compressed sensing, parallel imaging, and golden-angle radial
  sampling for fast and flexible dynamic volumetric mri,'' \emph{Magnetic
  resonance in medicine}, vol.~72, no.~3, pp. 707--717, 2014.

\bibitem{christodoulou2018magnetic}
A.~G. Christodoulou, J.~L. Shaw, C.~Nguyen, Q.~Yang, Y.~Xie, N.~Wang, and
  D.~Li, ``Magnetic resonance multitasking for motion-resolved quantitative
  cardiovascular imaging,'' \emph{Nature biomedical engineering}, vol.~2,
  no.~4, pp. 215--226, 2018.

\bibitem{feng2022golden}
L.~Feng, ``Golden-angle radial mri: basics, advances, and applications,''
  \emph{Journal of Magnetic Resonance Imaging}, vol.~56, no.~1, pp. 45--62,
  2022.

\bibitem{wallace2024rapid}
T.~E. Wallace, D.~Piccini, T.~Kober, S.~K. Warfield, and O.~Afacan, ``Rapid
  motion estimation and correction using self-encoded fid navigators in 3d
  radial mri,'' \emph{Magnetic resonance in medicine}, vol.~91, no.~3, pp.
  1057--1066, 2024.

\bibitem{schauman2026exploration}
S.~Schauman, A.~Van~Niekerk, O.~Norbeck, H.~Ryd{\'e}n, E.~Avventi, and
  S.~Skare, ``An exploration of motion-sampling interactions in 3d mri for
  neuroimaging,'' \emph{Magnetic Resonance in Medicine}, vol.~95, no.~3, pp.
  1448--1461, 2026.

\bibitem{noll1997multishot}
D.~C. Noll, ``Multishot rosette trajectories for spectrally selective mr
  imaging,'' \emph{IEEE transactions on medical imaging}, vol.~16, no.~4, pp.
  372--377, 1997.

\bibitem{ehman1984magnetic}
R.~L. Ehman, M.~McNamara, M.~Pallack, H.~Hricak, and C.~Higgins, ``Magnetic
  resonance imaging with respiratory gating: techniques and advantages,''
  \emph{American journal of Roentgenology}, vol. 143, no.~6, pp. 1175--1182,
  1984.

\bibitem{larson2004self}
A.~C. Larson, R.~D. White, G.~Laub, E.~R. McVeigh, D.~Li, and O.~P. Simonetti,
  ``Self-gated cardiac cine mri,'' \emph{Magnetic Resonance in Medicine: An
  Official Journal of the International Society for Magnetic Resonance in
  Medicine}, vol.~51, no.~1, pp. 93--102, 2004.

\bibitem{di2019automated}
L.~Di~Sopra, D.~Piccini, S.~Coppo, M.~Stuber, and J.~Yerly, ``An automated
  approach to fully self-gated free-running cardiac and respiratory
  motion-resolved 5d whole-heart mri,'' \emph{Magnetic resonance in medicine},
  vol.~82, no.~6, pp. 2118--2132, 2019.

\bibitem{ludwig2021pilot}
J.~Ludwig, P.~Speier, F.~Seifert, T.~Schaeffter, and C.~Kolbitsch, ``Pilot
  tone--based motion correction for prospective respiratory compensated cardiac
  cine mri,'' \emph{Magnetic Resonance in Medicine}, vol.~85, no.~5, pp.
  2403--2416, 2021.

\bibitem{anand2024beat}
S.~Anand and M.~Lustig, ``Beat pilot tone (bpt): Simultaneous mri and rf motion
  sensing at arbitrary frequencies,'' \emph{Magnetic resonance in medicine},
  vol.~92, no.~4, pp. 1768--1787, 2024.

\bibitem{moghari2013free}
M.~H. Moghari, S.~Roujol, R.~H. Chan, S.~N. Hong, N.~Bello, M.~Henningsson,
  L.~H. Ngo, B.~Goddu, L.~Goepfert, K.~V. Kissinger \emph{et~al.},
  ``Free-breathing 3d cardiac mri using iterative image-based respiratory
  motion correction,'' \emph{Magnetic Resonance in Medicine}, vol.~70, no.~4,
  pp. 1005--1015, 2013.

\bibitem{thesen2000prospective}
S.~Thesen, O.~Heid, E.~Mueller, and L.~R. Schad, ``Prospective acquisition
  correction for head motion with image-based tracking for real-time fmri,''
  \emph{Magnetic Resonance in Medicine: An Official Journal of the
  International Society for Magnetic Resonance in Medicine}, vol.~44, no.~3,
  pp. 457--465, 2000.

\bibitem{welch2002spherical}
E.~B. Welch, A.~Manduca, R.~C. Grimm, H.~A. Ward, and C.~R. Jack~Jr,
  ``Spherical navigator echoes for full 3d rigid body motion measurement in
  mri,'' \emph{Magnetic resonance in medicine: an official journal of the
  international society for magnetic resonance in medicine}, vol.~47, no.~1,
  pp. 32--41, 2002.

\bibitem{van2006real}
A.~J. Van~der Kouwe, T.~Benner, and A.~M. Dale, ``Real-time rigid body motion
  correction and shimming using cloverleaf navigators,'' \emph{Magnetic
  resonance in medicine: an official journal of the international society for
  magnetic resonance in medicine}, vol.~56, no.~5, pp. 1019--1032, 2006.

\bibitem{tisdall2012volumetric}
M.~D. Tisdall, A.~T. Hess, M.~Reuter, E.~M. Meintjes, B.~Fischl, and A.~J.
  van~der Kouwe, ``Volumetric navigators for prospective motion correction and
  selective reacquisition in neuroanatomical mri,'' \emph{Magnetic resonance in
  medicine}, vol.~68, no.~2, pp. 389--399, 2012.

\bibitem{white2010promo}
N.~White, C.~Roddey, A.~Shankaranarayanan, E.~Han, D.~Rettmann, J.~Santos,
  J.~Kuperman, and A.~Dale, ``Promo: real-time prospective motion correction in
  mri using image-based tracking,'' \emph{Magnetic Resonance in Medicine: An
  Official Journal of the International Society for Magnetic Resonance in
  Medicine}, vol.~63, no.~1, pp. 91--105, 2010.

\bibitem{zaitsev2017prospective}
M.~Zaitsev, B.~Akin, P.~LeVan, and B.~R. Knowles, ``Prospective motion
  correction in functional mri,'' \emph{Neuroimage}, vol. 154, pp. 33--42,
  2017.

\bibitem{slipsager2019markerless}
J.~M. Slipsager, A.~H. Ellegaard, S.~L. Glimberg, R.~R. Paulsen, M.~D. Tisdall,
  P.~Wighton, A.~Van Der~Kouwe, L.~Marner, O.~M. Henriksen, I.~Law
  \emph{et~al.}, ``Markerless motion tracking and correction for pet, mri, and
  simultaneous pet/mri,'' \emph{PloS one}, vol.~14, no.~4, p. e0215524, 2019.

\bibitem{slipsager2021comparison}
J.~Slipsager, S.~Glimberg, L.~H{\o}jgaard, R.~Paulsen, P.~Wighton, M.~Tisdall,
  C.~Jaimes, B.~Gagoski, P.~Grant, A.~van~der Kouwe \emph{et~al.}, ``Comparison
  of prospective and retrospective motion correction in 3d-encoded
  neuroanatomical mri.'' \emph{Magnetic Resonance in Medicine}, vol.~87, no.~2,
  pp. 629--645, 2021.

\bibitem{adanyeguh2024fast}
I.~M. Adanyeguh, N.~Bikkamane~Jayadev, P.-G. Henry, and D.~K. Deelchand, ``Fast
  high-resolution prospective motion correction for single-voxel
  spectroscopy,'' \emph{Magnetic resonance in medicine}, vol.~91, no.~4, pp.
  1301--1313, 2024.

\bibitem{jayadev2025accelerated}
N.~B. Jayadev, J.~Stockmann, R.~Frost, N.~Arango, Y.~Chang, A.~van~der Kouwe,
  and O.~C. Andronesi, ``Accelerated navigator for rapid {$\Delta B_0$} field
  mapping for real-time shimming and motion correction of human brain mri,''
  \emph{NMR in Biomedicine}, vol.~38, no.~10, p. e70126, 2025.

\bibitem{hansen2012retrospective}
M.~S. Hansen, T.~S. S{\o}rensen, A.~E. Arai, and P.~Kellman, ``Retrospective
  reconstruction of high temporal resolution cine images from real-time mri
  using iterative motion correction,'' \emph{Magnetic Resonance in Medicine},
  vol.~68, no.~3, pp. 741--750, 2012.

\bibitem{parkes2018evaluation}
L.~Parkes, B.~Fulcher, M.~Y{\"u}cel, and A.~Fornito, ``An evaluation of the
  efficacy, reliability, and sensitivity of motion correction strategies for
  resting-state functional mri,'' \emph{Neuroimage}, vol. 171, pp. 415--436,
  2018.

\bibitem{gallichan2016retrospective}
D.~Gallichan, J.~P. Marques, and R.~Gruetter, ``Retrospective correction of
  involuntary microscopic head movement using highly accelerated fat image
  navigators (3d fatnavs) at 7t,'' \emph{Magnetic resonance in medicine},
  vol.~75, no.~3, pp. 1030--1039, 2016.

\bibitem{marami2016motion}
B.~Marami, B.~Scherrer, O.~Afacan, B.~Erem, S.~K. Warfield, and A.~Gholipour,
  ``Motion-robust diffusion-weighted brain mri reconstruction through
  slice-level registration-based motion tracking,'' \emph{IEEE transactions on
  medical imaging}, vol.~35, no.~10, pp. 2258--2269, 2016.

\bibitem{godenschweger2016motion}
F.~Godenschweger, U.~K{\"a}gebein, D.~Stucht, U.~Yarach, A.~Sciarra,
  R.~Yakupov, F.~L{\"u}sebrink, P.~Schulze, and O.~Speck, ``Motion correction
  in mri of the brain,'' \emph{Physics in medicine \& biology}, vol.~61, no.~5,
  pp. R32--R56, 2016.

\bibitem{isaieva2025optimal}
K.~Isaieva, N.~Weber, L.~Nohava, B.~Fischer, A.~Megel, P.~Henrot, E.~Micard,
  E.~Laistler, J.~Felblinger, and F.~Odille, ``Optimal sensor selection for
  motion-corrected supine breast mri with a wearable coil,'' \emph{IEEE
  Transactions on Biomedical Engineering}, 2025.

\bibitem{aksoy2012hybrid}
M.~Aksoy, C.~Forman, M.~Straka, T.~{\c{C}}ukur, J.~Hornegger, and R.~Bammer,
  ``Hybrid prospective and retrospective head motion correction to mitigate
  cross-calibration errors,'' \emph{Magnetic resonance in medicine}, vol.~67,
  no.~5, pp. 1237--1251, 2012.

\bibitem{maclaren2011combined}
J.~Maclaren, K.~J. Lee, C.~Luengviriya, O.~Speck, and M.~Zaitsev, ``Combined
  prospective and retrospective motion correction to relax navigator
  requirements,'' \emph{Magnetic Resonance in Medicine}, vol.~65, no.~6, pp.
  1724--1732, 2011.

\bibitem{gholipour2011motion}
A.~Gholipour, M.~Polak, A.~Van Der~Kouwe, E.~Nevo, and S.~K. Warfield,
  ``Motion-robust mri through real-time motion tracking and retrospective
  super-resolution volume reconstruction,'' in \emph{2011 annual international
  conference of the IEEE engineering in medicine and biology society}.\hskip
  1em plus 0.5em minus 0.4em\relax IEEE, 2011, pp. 5722--5725.

\bibitem{andronesi2021motion}
O.~C. Andronesi, P.~K. Bhattacharyya, W.~Bogner, I.-Y. Choi, A.~T. Hess,
  P.~Lee, E.~M. Meintjes, M.~D. Tisdall, M.~Zaitsev, and A.~van~der Kouwe,
  ``Motion correction methods for mrs: experts' consensus recommendations,''
  \emph{NMR in Biomedicine}, vol.~34, no.~5, p. e4364, 2021.

\bibitem{olesen2012list}
O.~V. Olesen, J.~M. Sullivan, T.~Mulnix, R.~R. Paulsen, L.~Hojgaard, B.~Roed,
  R.~E. Carson, E.~D. Morris, and R.~Larsen, ``List-mode pet motion correction
  using markerless head tracking: proof-of-concept with scans of human
  subject,'' \emph{IEEE transactions on medical imaging}, vol.~32, no.~2, pp.
  200--209, 2012.

\bibitem{kober2011head}
T.~Kober, J.~P. Marques, R.~Gruetter, and G.~Krueger, ``Head motion detection
  using fid navigators,'' \emph{Magnetic resonance in medicine}, vol.~66,
  no.~1, pp. 135--143, 2011.

\bibitem{gholipour2010robust}
A.~Gholipour, J.~A. Estroff, and S.~K. Warfield, ``Robust super-resolution
  volume reconstruction from slice acquisitions: application to fetal brain
  mri,'' \emph{IEEE transactions on medical imaging}, vol.~29, no.~10, pp.
  1739--1758, 2010.

\bibitem{kainz2015fast}
B.~Kainz, M.~Steinberger, W.~Wein, M.~Kuklisova-Murgasova, C.~Malamateniou,
  K.~Keraudren, T.~Torsney-Weir, M.~Rutherford, P.~Aljabar, J.~V. Hajnal
  \emph{et~al.}, ``Fast volume reconstruction from motion corrupted stacks of
  2d slices,'' \emph{IEEE transactions on medical imaging}, vol.~34, no.~9, pp.
  1901--1913, 2015.

\bibitem{alansary2017pvr}
A.~Alansary, M.~Rajchl, S.~G. McDonagh, M.~Murgasova, M.~Damodaram, D.~F.
  Lloyd, A.~Davidson, M.~Rutherford, J.~V. Hajnal, D.~Rueckert \emph{et~al.},
  ``Pvr: patch-to-volume reconstruction for large area motion correction of
  fetal mri,'' \emph{IEEE transactions on medical imaging}, vol.~36, no.~10,
  pp. 2031--2044, 2017.

\bibitem{uus2020deformable}
A.~Uus, T.~Zhang, L.~H. Jackson, T.~A. Roberts, M.~A. Rutherford, J.~V. Hajnal,
  and M.~Deprez, ``Deformable slice-to-volume registration for motion
  correction of fetal body and placenta mri,'' \emph{IEEE transactions on
  medical imaging}, vol.~39, no.~9, pp. 2750--2759, 2020.

\bibitem{dosenbach2017real}
N.~U. Dosenbach, J.~M. Koller, E.~A. Earl, O.~Miranda-Dominguez, R.~L. Klein,
  A.~N. Van, A.~Z. Snyder, B.~J. Nagel, J.~T. Nigg, A.~L. Nguyen \emph{et~al.},
  ``Real-time motion analytics during brain mri improve data quality and reduce
  costs,'' \emph{Neuroimage}, vol. 161, pp. 80--93, 2017.

\bibitem{sui2020slimm}
Y.~Sui, O.~Afacan, A.~Gholipour, and S.~K. Warfield, ``Slimm: Slice
  localization integrated mri monitoring,'' \emph{NeuroImage}, vol. 223, p.
  117280, 2020.

\bibitem{marchetto2023robust}
E.~Marchetto, K.~Murphy, S.~L. Glimberg, and D.~Gallichan, ``Robust
  retrospective motion correction of head motion using navigator-based and
  markerless motion tracking techniques,'' \emph{Magnetic resonance in
  medicine}, vol.~90, no.~4, pp. 1297--1315, 2023.

\bibitem{neves2023real}
S.~Neves~Silva, J.~Aviles~Verdera, R.~Tomi-Tricot, R.~Neji, A.~Uus,
  I.~Grigorescu, T.~Wilkinson, V.~Ozenne, A.~Lewin, L.~Story \emph{et~al.},
  ``Real-time fetal brain tracking for functional fetal mri,'' \emph{Magnetic
  resonance in medicine}, vol.~90, no.~6, pp. 2306--2320, 2023.

\bibitem{verdera2025heron}
J.~A. Verdera, A.~Bortolazzi, S.~N. Silva, K.~Payette, K.~S. Clair, S.~McElroy,
  S.~Malik, J.~Hajnal, R.~Tomi-Tricot, M.~Rutherford \emph{et~al.}, ``Heron:
  High-efficiency real-time motion quantification and re-acquisition for fetal
  diffusion mri,'' \emph{IEEE Transactions on Medical Imaging}, 2025.

\bibitem{marchetto2026contrast}
E.~Marchetto, S.~Flassbeck, A.~Mao, and J.~Assl{\"a}nder, ``Contrast-optimized
  basis functions for self-navigated motion correction in quantitative mri,''
  \emph{Magnetic resonance in medicine}, vol.~95, no.~2, pp. 927--938, 2026.

\bibitem{fitzpatrick2000image}
J.~M. Fitzpatrick, D.~L. Hill, C.~R. Maurer \emph{et~al.}, ``Image
  registration,'' \emph{Handbook of medical imaging}, vol.~2, pp. 447--513,
  2000.

\bibitem{ferrante2017slice}
E.~Ferrante and N.~Paragios, ``Slice-to-volume medical image registration: A
  survey,'' \emph{Medical image analysis}, vol.~39, pp. 101--123, 2017.

\bibitem{gholipour2019biomedical}
A.~Gholipour and N.~Kehtarnavaz, ``Biomedical image registration,''
  \emph{Encyclopedia of Image Processing}, pp. 24--39, 2019.

\bibitem{salehi2018real}
S.~S.~M. Salehi, S.~Khan, D.~Erdogmus, and A.~Gholipour, ``Real-time deep pose
  estimation with geodesic loss for image-to-template rigid registration,''
  \emph{IEEE transactions on medical imaging}, vol.~38, no.~2, pp. 470--481,
  2018.

\bibitem{singh2020deep}
A.~Singh, S.~S.~M. Salehi, and A.~Gholipour, ``Deep predictive motion tracking
  in magnetic resonance imaging: application to fetal imaging,'' \emph{IEEE
  transactions on medical imaging}, vol.~39, no.~11, pp. 3523--3534, 2020.

\bibitem{calixto2024advances}
C.~Calixto, A.~Taymourtash, D.~Karimi, H.~Snoussi, C.~Velasco-Annis, C.~Jaimes,
  and A.~Gholipour, ``Advances in fetal brain imaging,'' \emph{Magnetic
  Resonance Imaging Clinics}, vol.~32, no.~3, pp. 459--478, 2024.

\bibitem{spieker2023deep}
V.~Spieker, H.~Eichhorn, K.~Hammernik, D.~Rueckert, C.~Preibisch, D.~C.
  Karampinos, and J.~A. Schnabel, ``Deep learning for retrospective motion
  correction in mri: a comprehensive review,'' \emph{IEEE Transactions on
  Medical Imaging}, 2023.

\bibitem{evan2022keymorph}
M.~Y. Evan, A.~Q. Wang, A.~V. Dalca, and M.~R. Sabuncu, ``Keymorph: Robust
  multi-modal affine registration via unsupervised keypoint detection,'' in
  \emph{International Conference on Medical Imaging with Deep Learning}.\hskip
  1em plus 0.5em minus 0.4em\relax PMLR, 2022, pp. 1482--1503.

\bibitem{wang2024joint}
J.~Wang, R.~Faghihpirayesh, D.~Erdogmus, and A.~Gholipour, ``Joint motion
  estimation with geometric deformation correction for fetal echo planar images
  via deep learning,'' in \emph{Medical Imaging with Deep Learning}, 2024.

\bibitem{wang2024spaer}
J.~Wang, R.~Faghihpirayesh, P.~Golland, and A.~Gholipour, ``Spaer: Learning
  spatio-temporal equivariant representations for fetal brain motion
  tracking,'' in \emph{International Workshop on Preterm, Perinatal and
  Paediatric Image Analysis}.\hskip 1em plus 0.5em minus 0.4em\relax Springer,
  2024, pp. 3--13.

\bibitem{moyer2021equivariant}
D.~Moyer, E.~Abaci~Turk, P.~E. Grant, W.~M. Wells, and P.~Golland,
  ``Equivariant filters for efficient tracking in 3d imaging,'' in
  \emph{Medical Image Computing and Computer Assisted Intervention--MICCAI
  2021: 24th International Conference, Strasbourg, France, September
  27--October 1, 2021, Proceedings, Part IV 24}.\hskip 1em plus 0.5em minus
  0.4em\relax Springer, 2021, pp. 193--202.

\bibitem{xu2022svort}
J.~Xu, D.~Moyer, P.~E. Grant, P.~Golland, J.~E. Iglesias, and E.~Adalsteinsson,
  ``Svort: iterative transformer for slice-to-volume registration in fetal
  brain mri,'' in \emph{International Conference on Medical Image Computing and
  Computer-Assisted Intervention}.\hskip 1em plus 0.5em minus 0.4em\relax
  Springer, 2022, pp. 3--13.

\bibitem{haskell2019network}
M.~W. Haskell, S.~F. Cauley, B.~Bilgic, J.~Hossbach, D.~N. Splitthoff,
  J.~Pfeuffer, K.~Setsompop, and L.~L. Wald, ``Network accelerated motion
  estimation and reduction (namer): convolutional neural network guided
  retrospective motion correction using a separable motion model,''
  \emph{Magnetic resonance in medicine}, vol.~82, no.~4, pp. 1452--1461, 2019.

\bibitem{schlemper2017deep}
J.~Schlemper, J.~Caballero, J.~V. Hajnal, A.~N. Price, and D.~Rueckert, ``A
  deep cascade of convolutional neural networks for dynamic mr image
  reconstruction,'' \emph{IEEE transactions on Medical Imaging}, vol.~37,
  no.~2, pp. 491--503, 2017.

\bibitem{lyu2021cine}
Q.~Lyu, H.~Shan, Y.~Xie, A.~C. Kwan, Y.~Otaki, K.~Kuronuma, D.~Li, and G.~Wang,
  ``Cine cardiac mri motion artifact reduction using a recurrent neural
  network,'' \emph{IEEE Transactions on Medical Imaging}, vol.~40, no.~8, pp.
  2170--2181, 2021.

\bibitem{xu2023nesvor}
J.~Xu, D.~Moyer, B.~Gagoski, J.~E. Iglesias, P.~E. Grant, P.~Golland, and
  E.~Adalsteinsson, ``Nesvor: implicit neural representation for
  slice-to-volume reconstruction in mri,'' \emph{IEEE transactions on medical
  imaging}, vol.~42, no.~6, pp. 1707--1719, 2023.

\bibitem{chatterjee2020retrospective}
S.~Chatterjee, A.~Sciarra, M.~D{\"u}nnwald, S.~Oeltze-Jafra, A.~N{\"u}rnberger,
  and O.~Speck, ``Retrospective motion correction of mr images using
  prior-assisted deep learning,'' \emph{arXiv preprint arXiv:2011.14134}, 2020.

\bibitem{al2023knowledge}
M.~A. Al-Masni, S.~Lee, A.~K. Al-Shamiri, S.-M. Gho, Y.~H. Choi, and D.-H. Kim,
  ``A knowledge interaction learning for multi-echo mri motion artifact
  correction towards better enhancement of swi,'' \emph{Computers in biology
  and medicine}, vol. 153, p. 106553, 2023.

\bibitem{kuzmina2022autofocusing+}
E.~Kuzmina, A.~Razumov, O.~Y. Rogov, E.~Adalsteinsson, J.~White, and D.~V.
  Dylov, ``Autofocusing+: noise-resilient motion correction in magnetic
  resonance imaging,'' in \emph{International Conference on Medical Image
  Computing and Computer-Assisted Intervention}.\hskip 1em plus 0.5em minus
  0.4em\relax Springer, 2022, pp. 365--375.

\bibitem{hossbach2023deep}
J.~Hossbach, D.~N. Splitthoff, S.~Cauley, B.~Clifford, D.~Polak, W.-C. Lo,
  H.~Meyer, and A.~Maier, ``Deep learning-based motion quantification from
  k-space for fast model-based magnetic resonance imaging motion correction,''
  \emph{Medical physics}, vol.~50, no.~4, pp. 2148--2161, 2023.

\bibitem{eichhorn2024physics}
H.~Eichhorn, V.~Spieker, K.~Hammernik, E.~Saks, K.~Weiss, C.~Preibisch, and
  J.~A. Schnabel, ``Physics-informed deep learning for motion-corrected
  reconstruction of quantitative brain mri,'' in \emph{MICCAI}, 2024.

\bibitem{singh2022joint}
N.~M. Singh, J.~E. Iglesias, E.~Adalsteinsson, A.~V. Dalca, and P.~Golland,
  ``Joint frequency and image space learning for mri reconstruction and
  analysis,'' \emph{The journal of machine learning for biomedical imaging},
  vol. 2022, 2022.

\bibitem{cui2023motion}
L.~Cui, Y.~Song, Y.~Wang, R.~Wang, D.~Wu, H.~Xie, J.~Li, and G.~Yang, ``Motion
  artifact reduction for magnetic resonance imaging with deep learning and
  k-space analysis,'' \emph{PloS one}, vol.~18, no.~1, p. e0278668, 2023.

\bibitem{seegoolam2019exploiting}
G.~Seegoolam, J.~Schlemper, C.~Qin, A.~Price, J.~Hajnal, and D.~Rueckert,
  ``Exploiting motion for deep learning reconstruction of
  extremely-undersampled dynamic mri,'' in \emph{International Conference on
  Medical Image Computing and Computer-Assisted Intervention}.\hskip 1em plus
  0.5em minus 0.4em\relax Springer, 2019, pp. 704--712.

\bibitem{qi2021end}
H.~Qi, R.~Hajhosseiny, G.~Cruz, T.~Kuestner, K.~Kunze, R.~Neji, R.~Botnar, and
  C.~Prieto, ``End-to-end deep learning nonrigid motion-corrected
  reconstruction for highly accelerated free-breathing coronary mra,''
  \emph{Magnetic Resonance in Medicine}, vol.~86, no.~4, pp. 1983--1996, 2021.

\bibitem{huang2023neural}
W.~Huang, H.~B. Li, J.~Pan, G.~Cruz, D.~Rueckert, and K.~Hammernik, ``Neural
  implicit k-space for binning-free non-cartesian cardiac mr imaging,'' in
  \emph{International Conference on Information Processing in Medical
  Imaging}.\hskip 1em plus 0.5em minus 0.4em\relax Springer, 2023, pp.
  548--560.

\bibitem{spieker2023iconik}
V.~Spieker, W.~Huang, H.~Eichhorn, J.~Stelter, K.~Weiss, V.~A. Zimmer, R.~F.
  Braren, D.~C. Karampinos, K.~Hammernik, and J.~A. Schnabel, ``Iconik:
  Generating respiratory-resolved abdominal mr reconstructions using neural
  implicit representations in k-space,'' in \emph{MICCAI}.\hskip 1em plus 0.5em
  minus 0.4em\relax Springer, 2023, pp. 183--192.

\bibitem{harput2018two}
S.~Harput, K.~Christensen-Jeffries, J.~Brown, Y.~Li, K.~J. Williams, A.~H.
  Davies, R.~J. Eckersley, C.~Dunsby, and M.-X. Tang, ``Two-stage motion
  correction for super-resolution ultrasound imaging in human lower limb,''
  \emph{IEEE transactions on ultrasonics, ferroelectrics, and frequency
  control}, vol.~65, no.~5, pp. 803--814, 2018.

\bibitem{prevost20183d}
R.~Prevost, M.~Salehi, S.~Jagoda, N.~Kumar, J.~Sprung, A.~Ladikos, R.~Bauer,
  O.~Zettinig, and W.~Wein, ``3d freehand ultrasound without external tracking
  using deep learning,'' \emph{Medical image analysis}, vol.~48, pp. 187--202,
  2018.

\bibitem{shi2021automatic}
L.~Shi, Y.~Lu, N.~Dvornek, C.~A. Weyman, E.~J. Miller, A.~J. Sinusas, and
  C.~Liu, ``Automatic inter-frame patient motion correction for dynamic cardiac
  pet using deep learning,'' \emph{IEEE transactions on medical imaging},
  vol.~40, no.~12, pp. 3293--3304, 2021.

\bibitem{li2021deep}
T.~Li, M.~Zhang, W.~Qi, E.~Asma, and J.~Qi, ``Deep learning based joint pet
  image reconstruction and motion estimation,'' \emph{IEEE transactions on
  medical imaging}, vol.~41, no.~5, pp. 1230--1241, 2021.

\bibitem{billot2024se}
B.~Billot, N.~Dey, D.~Moyer, M.~Hoffmann, E.~A. Turk, B.~Gagoski, P.~E. Grant,
  and P.~Golland, ``Se (3)-equivariant and noise-invariant 3d rigid motion
  tracking in brain mri,'' \emph{IEEE transactions on medical imaging},
  vol.~43, no.~11, pp. 4029--4040, 2024.

\bibitem{wang2023robust}
A.~Q. Wang, M.~Y. Evan, A.~V. Dalca, and M.~R. Sabuncu, ``A robust and
  interpretable deep learning framework for multi-modal registration via
  keypoints,'' \emph{Medical Image Analysis}, vol.~90, p. 102962, 2023.

\bibitem{wang2023metamorph}
J.~Wang, J.~Xing, J.~Druzgal, W.~M. Wells~III, and M.~Zhang, ``Metamorph:
  Learning metamorphic image transformation with appearance changes,'' in
  \emph{International Conference on Information Processing in Medical
  Imaging}.\hskip 1em plus 0.5em minus 0.4em\relax Springer, 2023, pp.
  576--587.

\bibitem{wang2024brainmorph}
A.~Q. Wang, R.~Saluja, H.~Kim, X.~He, A.~Dalca, and M.~R. Sabuncu,
  ``Brainmorph: A foundational keypoint model for robust and flexible brain mri
  registration,'' \emph{arXiv preprint arXiv:2405.14019}, 2024.

\bibitem{beg2005computing}
M.~F. Beg, M.~I. Miller, A.~Trouv{\'e}, and L.~Younes, ``Computing large
  deformation metric mappings via geodesic flows of diffeomorphisms,''
  \emph{International journal of computer vision}, vol.~61, pp. 139--157, 2005.

\bibitem{avants2008symmetric}
B.~B. Avants, C.~L. Epstein, M.~Grossman, and J.~C. Gee, ``Symmetric
  diffeomorphic image registration with cross-correlation: evaluating automated
  labeling of elderly and neurodegenerative brain,'' \emph{Medical image
  analysis}, vol.~12, no.~1, pp. 26--41, 2008.

\bibitem{wells1996multi}
W.~M. Wells~III, P.~Viola, H.~Atsumi, S.~Nakajima, and R.~Kikinis,
  ``Multi-modal volume registration by maximization of mutual information,''
  \emph{Medical image analysis}, vol.~1, no.~1, pp. 35--51, 1996.

\bibitem{arnold1966}
V.~I. Arnol'd, ``Sur la g\'{e}om\'{e}trie diff\'{e}rentielle des groupes de
  {L}ie de dimension infinie et ses applications \`{a} l'hydrodynamique des
  fluides parfaits,'' \emph{Ann. Inst. Fourier}, vol.~16, pp. 319--361, 1966.

\bibitem{miller2006geodesic}
M.~I. Miller, A.~Trouv{\'e}, and L.~Younes, ``Geodesic shooting for
  computational anatomy,'' \emph{Journal of mathematical imaging and vision},
  vol.~24, no.~2, pp. 209--228, 2006.

\bibitem{zhang2019fast}
M.~Zhang and P.~T. Fletcher, ``Fast diffeomorphic image registration via
  fourier-approximated lie algebras,'' \emph{International Journal of Computer
  Vision}, vol. 127, no.~1, pp. 61--73, 2019.

\bibitem{faghihpirayesh2024fetal}
R.~Faghihpirayesh, D.~Karimi, D.~Erdo{\u{g}}mu{\c{s}}, and A.~Gholipour,
  ``Fetal-bet: Brain extraction tool for fetal mri,'' \emph{IEEE Open Journal
  of Engineering in Medicine and Biology}, 2024.

\bibitem{vranic20013d}
D.~Vranic and D.~Saupe, ``3d shape descriptor based on 3d fourier transform,''
  2001.

\bibitem{khotanzad1990invariant}
A.~Khotanzad and Y.~H. Hong, ``Invariant image recognition by zernike
  moments,'' \emph{IEEE Transactions on pattern analysis and machine
  intelligence}, vol.~12, no.~5, pp. 489--497, 1990.

\bibitem{jafari2014spherical}
M.~Jafari and H.~Molaei, ``Spherical linear interpolation and b{\'e}zier
  curves,'' \emph{General Scientific Researches}, vol.~2, no.~1, pp. 13--17,
  2014.

\bibitem{zhong2024slerpface}
Z.~Zhong, Y.~Mi, Y.~Huang, J.~Xu, G.~Mu, S.~Ding, J.~Zhang, R.~Guo, Y.~Wu, and
  S.~Zhou, ``Slerpface: Face template protection via spherical linear
  interpolation,'' \emph{arXiv preprint arXiv:2407.03043}, 2024.

\bibitem{jang2024spherical}
Y.~K. Jang, D.~Huynh, A.~Shah, W.-K. Chen, and S.-N. Lim, ``Spherical linear
  interpolation and text-anchoring for zero-shot composed image retrieval,''
  \emph{arXiv preprint arXiv:2405.00571}, 2024.

\bibitem{li2022quatse}
J.~Li, X.~Su, X.~Ma, and G.~Gao, ``Quatse: Spherical linear interpolation of
  quaternion for knowledge graph embeddings,'' in \emph{CCF International
  Conference on Natural Language Processing and Chinese Computing}.\hskip 1em
  plus 0.5em minus 0.4em\relax Springer, 2022, pp. 209--220.

\bibitem{balakrishnan2019voxelmorph}
G.~Balakrishnan, A.~Zhao, M.~R. Sabuncu, J.~Guttag, and A.~V. Dalca,
  ``Voxelmorph: a learning framework for deformable medical image
  registration,'' \emph{IEEE transactions on medical imaging}, vol.~38, no.~8,
  pp. 1788--1800, 2019.

\bibitem{chen2022transmorph}
J.~Chen, E.~C. Frey, Y.~He, W.~P. Segars, Y.~Li, and Y.~Du, ``Transmorph:
  Transformer for unsupervised medical image registration,'' \emph{Medical
  image analysis}, vol.~82, p. 102615, 2022.

\bibitem{kim2022diffusemorph}
B.~Kim, I.~Han, and J.~C. Ye, ``Diffusemorph: Unsupervised deformable image
  registration using diffusion model,'' in \emph{European conference on
  computer vision}.\hskip 1em plus 0.5em minus 0.4em\relax Springer, 2022, pp.
  347--364.

\bibitem{nocedal1999numerical}
J.~Nocedal and S.~J. Wright, \emph{Numerical optimization}.\hskip 1em plus
  0.5em minus 0.4em\relax Springer, 1999.

\bibitem{yang2017data}
\BIBentryALTinterwordspacing
J.~Yang, G.~Sharp, H.~Veeraraghavan, W.~Van~Elmpt, A.~Dekker, T.~Lustberg, and
  M.~Gooding, ``Data from lung ct segmentation challenge (lctsc)(version
  3)[data set],'' \emph{The Cancer Imaging Archive}, 2017. [Online]. Available:
  \url{https://doi.org/10.7937/K9/TCIA.2017.3R3FVZ08}
\BIBentrySTDinterwordspacing

\bibitem{yang2021medmnist}
J.~Yang, R.~Shi, and B.~Ni, ``Medmnist classification decathlon: A lightweight
  automl benchmark for medical image analysis,'' in \emph{2021 IEEE 18th
  International Symposium on Biomedical Imaging (ISBI)}.\hskip 1em plus 0.5em
  minus 0.4em\relax IEEE, 2021, pp. 191--195.

\bibitem{medmnistv2}
J.~Yang, R.~Shi, D.~Wei, Z.~Liu, L.~Zhao, B.~Ke, H.~Pfister, and B.~Ni,
  ``Medmnist v2-a large-scale lightweight benchmark for 2d and 3d biomedical
  image classification,'' \emph{Scientific Data}, vol.~10, no.~1, p.~41, 2023.

\bibitem{baid2023}
\BIBentryALTinterwordspacing
U.~Baid, S.~Ghodasara, S.~Mohan, M.~Bilello, E.~Calabrese, E.~Colak,
  K.~Farahani, J.~Kalpathy-Cramer, F.~C. Kitamura, S.~Pati, L.~Prevedello,
  J.~Rudie, C.~Sako, R.~Shinohara, T.~Bergquist, R.~Chai, J.~Eddy, J.~Elliott,
  W.~Reade, T.~Schaffter, T.~Yu, J.~Zheng, C.~Davatzikos, J.~Mongan, C.~Hess,
  S.~Cha, J.~Villanueva-Meyer, J.~B. Freymann, J.~S. Kirby, B.~Wiestler,
  P.~Crivellaro, R.~R. Colen, A.~Kotrotsou, D.~Marcus, M.~Milchenko, A.~Nazeri,
  H.~Fathallah-Shaykh, R.~Wiest, A.~Jakab, M.-A. Weber, A.~Mahajan, B.~Menze,
  A.~E. Flanders, and S.~Bakas, ``Rsna-asnr-miccai-brats-2021 dataset,''
  \emph{The Cancer Imaging Archive}, 2023. [Online]. Available:
  \url{https://doi.org/10.7937/jc8x-9874}
\BIBentrySTDinterwordspacing

\bibitem{baid2021rsna}
U.~Baid, S.~Ghodasara, S.~Mohan, M.~Bilello, and E.~Calabrese, ``The
  rsna-asnr-miccai brats 2021 benchmark on brain tumor segmentation and
  radiogenomic classification,'' \emph{arXiv preprint arXiv:2107.02314}, 2021.

\bibitem{menze2014multimodal}
B.~H. Menze, A.~Jakab, S.~Bauer, J.~Kalpathy-Cramer, K.~Farahani, J.~Kirby,
  Y.~Burren, N.~Porz, J.~Slotboom, R.~Wiest \emph{et~al.}, ``The multimodal
  brain tumor image segmentation benchmark (brats),'' \emph{IEEE transactions
  on medical imaging}, vol.~34, no.~10, pp. 1993--2024, 2014.

\bibitem{afacan2019fetal}
O.~Afacan, J.~A. Estroff, E.~Yang, C.~E. Barnewolt, S.~A. Connolly, R.~B.
  Parad, R.~V. Mulkern, S.~K. Warfield, and A.~Gholipour, ``Fetal echoplanar
  imaging: promises and challenges,'' \emph{Topics in Magnetic Resonance
  Imaging}, vol.~28, no.~5, pp. 245--254, 2019.

\bibitem{christiaens2019utero}
D.~Christiaens, P.~J. Slator, L.~Cordero-Grande, A.~N. Price, M.~Deprez, D.~C.
  Alexander, M.~Rutherford, J.~V. Hajnal, and J.~Hutter, ``In utero diffusion
  mri: challenges, advances, and applications,'' \emph{Topics in Magnetic
  Resonance Imaging}, vol.~28, no.~5, pp. 255--264, 2019.

\bibitem{pier20163d}
D.~B. Pier, A.~Gholipour, O.~Afacan, C.~Velasco-Annis, S.~Clancy, K.~Kapur,
  J.~A. Estroff, and S.~K. Warfield, ``3d super-resolution motion-corrected
  mri: validation of fetal posterior fossa measurements,'' \emph{Journal of
  Neuroimaging}, vol.~26, no.~5, pp. 539--544, 2016.

\bibitem{roy2019fetal}
C.~W. Roy, J.~F. van Amerom, D.~Marini, M.~Seed, and C.~K. Macgowan, ``Fetal
  cardiac mri: a review of technical advancements,'' \emph{Topics in magnetic
  resonance imaging}, vol.~28, no.~5, pp. 235--244, 2019.

\bibitem{van2019fetal}
J.~F. van Amerom, D.~F. Lloyd, M.~Deprez, A.~N. Price, S.~J. Malik,
  K.~Pushparajah, M.~P. van Poppel, M.~A. Rutherford, R.~Razavi, and J.~V.
  Hajnal, ``Fetal whole-heart 4d imaging using motion-corrected multi-planar
  real-time mri,'' \emph{Magnetic resonance in medicine}, vol.~82, no.~3, pp.
  1055--1072, 2019.

\bibitem{vollbrecht2024fetal}
T.~M. Vollbrecht, M.~M. Bissell, F.~Kording, A.~Geipel, A.~Isaak, B.~S.
  Strizek, C.~Hart, A.~J. Barker, and J.~A. Luetkens, ``Fetal cardiac mri using
  doppler us gating: emerging technology and clinical implications,''
  \emph{Radiology: Cardiothoracic Imaging}, vol.~6, no.~2, p. e230182, 2024.

\bibitem{vollbrecht2025improving}
T.~Vollbrecht, C.~Hart, C.~Katemann, A.~Isaak, M.~Voigt, C.~Pieper,
  D.~Kuetting, A.~Geipel, B.~Strizek, and J.~Luetkens, ``Improving clinical
  utility of fetal cine cmr using deep learning super-resolution.''
  \emph{Circulation. Cardiovascular Imaging}, vol.~18, no.~8, pp.
  e018\,090--e018\,090, 2025.

\bibitem{roberts2020fetal}
T.~A. Roberts, J.~F. van Amerom, A.~Uus, D.~F. Lloyd, M.~P. van Poppel, A.~N.
  Price, J.-D. Tournier, C.~A. Mohanadass, L.~H. Jackson, S.~J. Malik
  \emph{et~al.}, ``Fetal whole heart blood flow imaging using 4d cine mri,''
  \emph{Nature communications}, vol.~11, no.~1, p. 4992, 2020.

\bibitem{van2023fetal}
J.~F. van Amerom, D.~S. Goolaub, E.~M. Schrauben, L.~Sun, C.~K. Macgowan, and
  M.~Seed, ``Fetal cardiovascular blood flow mri: techniques and
  applications,'' \emph{The British Journal of Radiology}, vol.~96, no. 1147,
  p. 20211096, 2023.

\bibitem{tompkins2025third}
R.~M. Tompkins, T.~Fujiwara, E.~M. Schrauben, L.~P. Browne, J.~van Schuppen,
  S.-A. Clur, R.~M. Friesen, E.~K. Englund, A.~J. Barker, and P.~van Ooij,
  ``Third trimester fetal 4d flow mri with motion correction,'' \emph{Magnetic
  Resonance in Medicine}, vol.~93, no.~5, pp. 1969--1983, 2025.

\bibitem{marami2017temporal}
B.~Marami, S.~S.~M. Salehi, O.~Afacan, B.~Scherrer, C.~K. Rollins, E.~Yang,
  J.~A. Estroff, S.~K. Warfield, and A.~Gholipour, ``Temporal slice
  registration and robust diffusion-tensor reconstruction for improved fetal
  brain structural connectivity analysis,'' \emph{NeuroImage}, vol. 156, pp.
  475--488, 2017.

\bibitem{snoussi2025haitch}
H.~Snoussi, D.~Karimi, O.~Afacan, M.~Utkur, and A.~Gholipour, ``Haitch: A
  framework for distortion and motion correction in fetal multi-shell
  diffusion-weighted mri,'' \emph{Imaging Neuroscience}, vol.~3, p.
  imag\_a\_00490, 2025.

\bibitem{sobotka2022motion}
D.~Sobotka, M.~Ebner, E.~Schwartz, K.-H. Nenning, A.~Taymourtash,
  T.~Vercauteren, S.~Ourselin, G.~Kasprian, D.~Prayer, G.~Langs \emph{et~al.},
  ``Motion correction and volumetric reconstruction for fetal functional
  magnetic resonance imaging data,'' \emph{NeuroImage}, vol. 255, p. 119213,
  2022.

\bibitem{taymourtash2025measuring}
A.~Taymourtash, E.~Schwartz, K.-H. Nenning, R.~Licandro, P.~Kienast, V.~Hielle,
  D.~Prayer, G.~Kasprian, and G.~Langs, ``Measuring the effects of motion
  corruption in fetal fmri,'' \emph{Human Brain Mapping}, vol.~46, no.~2, p.
  e26806, 2025.

\bibitem{benner2011diffusion}
T.~Benner, A.~J. van~der Kouwe, and A.~G. Sorensen, ``Diffusion imaging with
  prospective motion correction and reacquisition,'' \emph{Magnetic resonance
  in medicine}, vol.~66, no.~1, pp. 154--167, 2011.

\bibitem{gagoski2022automated}
B.~Gagoski, J.~Xu, P.~Wighton, M.~D. Tisdall, R.~Frost, W.-C. Lo, P.~Golland,
  A.~van Der~Kouwe, E.~Adalsteinsson, and P.~E. Grant, ``Automated detection
  and reacquisition of motion-degraded images in fetal haste imaging at 3 t,''
  \emph{Magnetic resonance in medicine}, vol.~87, no.~4, pp. 1914--1922, 2022.

\bibitem{avants2011reproducible}
B.~B. Avants, N.~J. Tustison, G.~Song, P.~A. Cook, A.~Klein, and J.~C. Gee, ``A
  reproducible evaluation of ants similarity metric performance in brain image
  registration,'' \emph{Neuroimage}, vol.~54, no.~3, pp. 2033--2044, 2011.

\bibitem{jian2010robust}
B.~Jian and B.~C. Vemuri, ``Robust point set registration using gaussian
  mixture models,'' \emph{IEEE transactions on pattern analysis and machine
  intelligence}, vol.~33, no.~8, pp. 1633--1645, 2010.

\bibitem{wasza2012real}
J.~Wasza, S.~Bauer, and J.~Hornegger, ``Real-time motion compensated patient
  positioning and non-rigid deformation estimation using 4-d shape priors,'' in
  \emph{International Conference on Medical Image Computing and
  Computer-Assisted Intervention}.\hskip 1em plus 0.5em minus 0.4em\relax
  Springer, 2012, pp. 576--583.

\bibitem{qi2020non}
H.~Qi, N.~Fuin, G.~Cruz, J.~Pan, T.~Kuestner, A.~Bustin, R.~M. Botnar, and
  C.~Prieto, ``Non-rigid respiratory motion estimation of whole-heart coronary
  mr images using unsupervised deep learning,'' \emph{IEEE Transactions on
  Medical Imaging}, vol.~40, no.~1, pp. 444--454, 2020.

\bibitem{rong2021rigid}
Y.~Rong, M.~Rosu-Bubulac, S.~H. Benedict, Y.~Cui, R.~Ruo, T.~Connell,
  R.~Kashani, K.~Latifi, Q.~Chen, H.~Geng \emph{et~al.}, ``Rigid and deformable
  image registration for radiation therapy: a self-study evaluation guide for
  nrg oncology clinical trial participation,'' \emph{Practical radiation
  oncology}, vol.~11, no.~4, pp. 282--298, 2021.

\bibitem{lv2022joint}
J.~Lv, Z.~Wang, H.~Shi, H.~Zhang, S.~Wang, Y.~Wang, and Q.~Li, ``Joint
  progressive and coarse-to-fine registration of brain mri via deformation
  field integration and non-rigid feature fusion,'' \emph{IEEE Transactions on
  Medical Imaging}, vol.~41, no.~10, pp. 2788--2802, 2022.

\bibitem{kurugol2017motion}
S.~Kurugol, M.~Freiman, O.~Afacan, L.~Domachevsky, J.~M. Perez-Rossello, M.~J.
  Callahan, and S.~K. Warfield, ``Motion-robust parameter estimation in
  abdominal diffusion-weighted mri by simultaneous image registration and model
  estimation,'' \emph{Medical image analysis}, vol.~39, pp. 124--132, 2017.

\bibitem{timms2026fast}
L.~Timms, M.~Utkur, C.~Ariyurek, M.~Hewlett, S.~Kurugol, and O.~Afacan, ``Fast,
  robust t2-ivim quantitative mri with distortion and motion-corrected
  multi-echo epi,'' \emph{Magnetic Resonance in Medicine}, 2026.

\bibitem{hinkle2018lagomorph}
J.~D. Hinkle, ``Lagomorph,'' Oak Ridge National Laboratory (ORNL), Oak Ridge,
  TN (United States), Tech. Rep., 2018.

\end{thebibliography}

\begin{IEEEbiography}[{\includegraphics[width=1in,height=1.25in,clip,keepaspectratio]{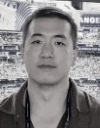}}]{Jian Wang} received the Ph.D. degree in Computer Science from the University of Virginia, Charlottesville, VA, USA, in 2023. He is currently a Research Advisor at Eli Lilly and Company, Indianapolis, IN, USA. Prior to his current role, he served as a Postdoctoral Research Fellow at Boston Children's Hospital and Harvard Medical School, Boston, MA, USA. His research interests lie at the intersection of machine learning, computer vision and medical image analysis, with a focus on deep learning methodologies, image registration, motion correction, shape analysis, and generative models.
\end{IEEEbiography}

\begin{IEEEbiography}[{\includegraphics[width=1in,height=1.25in,clip,keepaspectratio]{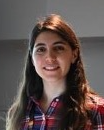}}]{Razieh Faghihpirayesh} received the Ph.D. degree in Electrical Engineering from Northeastern University, Boston, MA, USA, in 2025. She is currently an AI/ML Postdoctoral Researcher with Johnson and Johnson Innovative Medicine, where she develops artificial intelligence (AI) and machine learning (ML) methods for healthcare applications. Her research interests include medical image analysis, computer vision, and deep learning.
\end{IEEEbiography}

\begin{IEEEbiography}[{\includegraphics[width=1in,height=1.25in,clip,keepaspectratio]{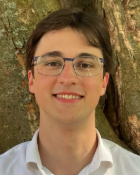}}]{Danny Joca} received his B.S. in Computer Engineering from Brown University, Providence, RI, USA, in 2023. He is currently a Research Assistant at Boston Children's Hospital, Boston, MA, USA, with the Musculoskeletal Digital Innovation and Informatics Program (MDI²), where he works on a full 3D hip assessment pipeline (VirtualHip). He was previously with the Computational Radiology Laboratory at Boston Children's Hospital, where his work centered on fetal brain MRI, including prospective image registration and deep learning-based image segmentation pipelines. His research interests include image processing and machine learning methods for medical image analysis.
\end{IEEEbiography}

\begin{IEEEbiography}[{\includegraphics[width=1in,height=1.25in,clip,keepaspectratio]{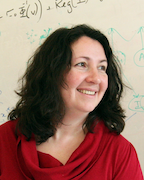}}]{Polina Golland} is a Sunlin (1966) and Priscilla Chou professor of
Electrical Engineering and Computer Science at the Massachusetts Institute of Technology (MIT), Cambridge, MA, USA, and a principal
investigator in the MIT Computer Science and Artificial Intelligence
Laboratory (CSAIL). Her primary research interest is in developing
novel techniques for medical image analysis and understanding. With
her students, Polina has demonstrated novel approaches to image
segmentation, shape analysis, functional image analysis and population
studies. She has served as an associate editor of the IEEE
Transactions on Medical Imaging and of the IEEE Transactions on
Pattern Analysis. Polina is currently on the editorial board of the
Journal of Medical Image Analysis. She is a Fellow of the
International Society for Medical Image Computing and Computer
Assisted Interventions (MICCAI) and of the American Institute for
Medical and Biological Engineering (AIMBE).
\end{IEEEbiography}

\begin{IEEEbiography}[{\includegraphics[width=1in,height=1.25in,clip,keepaspectratio]{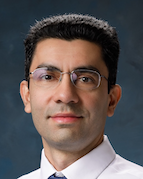}}]{Ali Gholipour} (M’08–SM’15) received the B.S. and M.S. degrees in electrical engineering from the University of Tehran, Tehran, Iran, in 2001 and 2003, respectively, and the Ph.D. degree in electrical engineering from the University of Texas at Dallas, Richardson, TX, USA, in 2008. He is currently a Professor of Radiological Sciences, Electrical Engineering and Computer Science, and Biomedical Engineering at the University of California, Irvine (UCI), CA, USA. Prior to this, from 2010 to 2024, he was a faculty at Boston Children’s Hospital, Harvard Medical School, Boston, MA, USA. He is serving as an Associate Editor for the IEEE Transactions on Medical Imaging as well as the journal of Medical Image Analysis. His primary research interest lies in the broad area of machine learning in medical imaging and image computing.
\end{IEEEbiography}

%If you do not have or do not want to include a photo, you can use IEEEbiographynophoto as shown below:

\EOD

\end{document}